\documentclass[11pt,a4]{article}                               

\usepackage{latexsym}
\usepackage{amssymb}
\usepackage{mathrsfs}
\usepackage{amsmath}
\usepackage{amsthm}
\usepackage{mathtools}
\usepackage{color}
\usepackage{graphicx}
\usepackage{tikz}
\usepackage{jheppub}

\newcommand{\nc}{\newcommand}
\newcommand{\rnc}{\renewcommand}

\nc{\be}{\begin{equation}}
\nc{\ee}{\end{equation}}
\nc{\bea}{\begin{align}}
\nc{\eea}{\end{align}}
\nc{\ben}{\begin{enumerate}}
\nc{\een}{\end{enumerate}}

\nc{\del}{\partial}
\nc{\la}{\lambda}
\nc{\eps}{\epsilon}

\nc{\CC}{\mathbb{C}}
\nc{\EE}{\mathbb{E}}
\nc{\NN}{\mathbb{N}}
\nc{\QQ}{\mathbb{Q}}
\nc{\RR}{\mathbb{R}}
\nc{\ZZ}{\mathbb{Z}}

\nc{\trac}[2]{{\textstyle\frac{#1}{#2}}}

\nc{\ex}[1]{\mbox{e}^{\,\textstyle#1}}

\nc{\tr}{\mathop{\mathrm{tr}}\nolimits}
\nc{\Ad}{\mathsf{Ad}}
\nc{\ad}{\mathsf{ad}}
\nc{\Tr}{\mathop{\mathrm{Tr}}\nolimits}
\nc{\Det}{\mathop{\mathrm{Det}}\nolimits}
\nc{\rk}{\mathop{\mathrm{rk}}\nolimits}
\nc{\ra}{\rightarrow}
\nc{\Ra}{\Rightarrow}
\nc{\LRa}{\Leftrightarrow}
\nc{\ot}{\otimes}
\nc{\nul}{\noindent\underline}
\nc{\non}{\nonumber\\}

\nc{\lf}{\mathfrak{f}}
\rnc{\lg}{\mathfrak{g}}
\nc{\lk}{\mathfrak{k}}
\nc{\lkk}{\mathfrak{kk}}
\nc{\lm}{\mathfrak{m}}
\nc{\lp}{\mathfrak{p}}
\nc{\ls}{\mathfrak{s}}
\nc{\lt}{\mathfrak{t}}
\nc{\ltg}{\mathfrak{tg}}
\nc{\lttg}{\mathfrak{ttg}}
\nc{\ltt}{\mathfrak{tt}}

\title{3 Definitions of BF Theory on Homology 3-Spheres}
\author[a]{Matthias Blau}
\author[b,c]{Mbambu Kakona}
\author[c]{George Thompson}

\affiliation[a]{Albert Einstein Center for Fundamental Physics\\ 
Institute for
Theoretical Physics, University of Bern, Switzerland}
\affiliation[b]{East African Institute for Fundamental Research (EAIFR)\\
University of Rwanda, Kigali, Rwanda}
\affiliation[c]{Abdus Salam International Centre for Theoretical Physics\\
Strada Costiera 11 Trieste, Italy}

\emailAdd{blau@itp.unibe.ch}
\emailAdd{kakona@eaifr.org}
\emailAdd{thompson@ictp.it}
\abstract{3-dimensional BF theory with gauge group $G$ (= Chern-Simons theory
with non-compact gauge group $TG$) is a deceptively simple yet
subtle topological gauge theory. Formally, its partition function
is a sum/integral over the moduli space $\mathcal{M}$ of flat
connections, weighted by the Ray-Singer torsion. In practice,
however, this formal expression is almost invariably singular and
ill-defined.

In order to improve upon this, we perform a direct evaluation of
the path integral for certain classes of 3-manifolds (namely integral
and rational Seifert homology spheres). By a suitable choice of
gauge, we sidestep the issue of having to integrate over $\mathcal{M}$
and reduce the partition function to a finite-dimensional Abelian
matrix integral which, however, itself requires a definition. We
offer 3 definitions of this integral, firstly via residues, and
then via a large $k$ limit of the corresponding $G\times G$ or
$G_\CC$ Chern-Simons matrix integrals (obtained previously). We
then check and discuss to which extent the results capture the
expected sum/integral over all flat connections.
  }

\global\parskip=4pt

\begin{document}
\maketitle

\section{Introduction}

Gauge theories with non-compact gauge groups are notoriously difficult
to make sense of, issues including questions of convergence and 
propagators with the wrong
signs which make unitarity and hence the physical meaning of
the theory far from clear. One set of theories for which one may make
some headway are topological field theories where a particle
interpretation is not required. Our first investigation in that
direction where the partition function can be evaluated exactly
\cite{BKT-MRST} was to consider the Schwarz type theories 
\cite{Schwarz-Partition, Schwarz, Schwarz-Tyupkin} formally representing 
the Ray-Singer Torsion 
\cite{Ray-Singer,Ray-Singer-Analytic} 
which
have non-compact symmetries. These symmetries though are essentially
Abelian and we would like to consider now a class of theories with non-Abelian
non-compact gauge groups. In particular we have in mind the
topological $BF$ theories \cite{BT-RST, Horowitz, BT-BF, BT-Metric} based on a compact gauge
group $G$. In two dimensions
such a theory is the zero coupling limit of Yang-Mills theory with gauge group $G$ and as
such does not have an associated non-compact symmetry. In
dimension greater than or equal to 3, however, $BF$ theories possess additional 
non-compact shift symmetries. We will focus on
3-dimensional $BF$ theory here, where the combined gauge group turns out to be the non-compact
group $TG$, the tangent bundle group of $G$.

Witten \cite{WBF} first introduced 3-dimensional $BF$ theories as a variant
of Chern-Simons theory \cite{Witten-CS} for non-compact gauge groups (in fact, it 
can be regarded as a Chern-Simons theory for the non-compact gauge group $TG$). The action is
\be
I_{BF}= \int_{M} \Tr{\left( B \wedge F_{A}\right)} \label{BF}
\ee
where $A$ is a connection on a $G$-bundle over $M$ and $B$ is a Lie
algebra valued 1-form. The action enjoys the usual compact $G$ gauge symmetry, 
as well as a non-compact shift gauge symmetry $B \ra B+ d_{A}\Lambda$.
This theory is deceptively simple. The path integral over $B$
yields a delta function constraint on the curvature 2-form $F_{A}$, 
so that formally the path integral for the partition fuction of $BF$ theory is
\begin{equation}
Z_{BF}[M,G] = \int DADB\, \exp{\left(i\int_{M}
  \Tr{\left( B \wedge F_{A
              }\right)}\right)}
  \simeq \int DA \delta\left(F_{A} \right)
\end{equation}
This seems to suggest that the path integral simply reduces to an 
integral over the moduli space of flat
connections, with some measure. In general, however, this is not correct. 
Rather, the complete classical equations of motion are
\be
F_{A}=0, \;\;\; d_{A}B=0\label{eq-motion}
\ee
which taken together (and modulo gauge transformations) 
can formally be viewed as describing the tangent bundle of
the moduli space $\mathfrak{M}[M,G]$ of flat $G$ connections on the
3-manifold $M$, $T\mathfrak{M}[M,G]$.
The partition function is then an integral over this space
(and ghost zero modes etc.), again with some measure to be determined.
However, in general this space can be very singular and 
the finite dimensional integral is ill-defined. The various
types of singularities that lurk inside the path integral associated
with the action (\ref{BF}) include issues with reducible connections
(manifested as ghost zero modes) as well as the non-compactness of
the moduli space itself (due to $B$ zero modes).

One situation in which it is pretty clear what sort of contribution
one should find from the path integral is when one expands the path
integral around an isolated irreducible flat connection. In that
case, there are no non-trivial solutions to the equations of motion
for $B$ and no ghost zero modes, and following Witten
\cite{WBF}, the contribution of such an \textit{acyclic} connection
$\omega$ on a 3-manifold $M$ can be shown to be precisely the
Ray-Singer analytic torsion $\tau_M(\omega)$ 
\cite{Ray-Singer,Ray-Singer-Analytic} 
of that connection.
Thus on manifolds on which the moduli space of flat connections 
includes such acyclic flat connections, one should
perhaps expect the partition function to take the form
\be
Z_{BF}[M,G] = \sum_{\mathrm{acyclic}} \tau_{M}(\omega_{(\alpha)})
 + \ldots 
\label{IZ-plus}
\ee
where the ellipses indicate the contributions  from the non-acyclic 
flat connections. If $M$ is an Integral Homology Sphere ($\mathbb{Z}$ Homology Sphere,
$\mathrm{H}_{1}(M,\mathbb{Z})=0$), then for $G=SU(2)$ the trivial
connection is the only reducible flat connection, and there are indeed
$\mathbb{Z}$ Homology Spheres for which all the flat connections are isolated.
In such a situation one should expect the $BF$ partition function to be largely
captured by the above expression. One of the aims of this paper will be to 
check to which extent this expectation is borne out by actual explicit evaluations
of the path integral of $BF$ theory.

Moreover, by using the Massive Ray-Singer Torsion $\tau_M(\omega,\mu)$ introduced in \cite{BKT-MRST}
(among other reasons precisely with its application to $BF$ theory in mind),
one can extend the considerations of Witten to the case that the
flat connections are not necessarily irreducible and isolated, and thus
obtain a formal expression for the regularised partition function of $BF$ theory
(section \ref{section-2}), which takes the form \eqref{BF-implicit}
\be
Z_{BF}[M,G;\mu] = \int\mathllap{\sum} \, \tau_{M}(\omega_{(
          \alpha)}, \mu) \label{BF-implicit}
\ee
of a sum / integral over the moduli space of flat connections. This expression
can in turn be expanded in the mass-parameter $\mu$ to reproduce different kinds
of contributions to the partition function \cite{BKT-MRST}. 
However, this approach outlined above to defining the partition function of
$BF$ theory is rather formal: one would need to determine by hand
the spaces of flat connections and zero modes associated with them,
calculate the (massive) Ray-Singer torsion by some means, etc. So
this does not all by itself lead to an \textit{evaluation} of the $BF$ partition function
via path integral methods.

In order to be able to verify to which extent an actual evaluation
of the $BF$ path integral reproduces the above formal expectations,
one needs a class of manifolds for which explicit path integral calculations
can be performed. To that end
we will concentrate on Seifert fibred 3-manifolds. These spaces
allow for a significant simplification in the calculations via suitable gauge
choices. The first of these is that the `time' (fibre) components
of both $A$ and $B$ (the components of $A$ and $B$ along the fibres of the defining 
Seifert fibrations) 
can be taken to be constant along the fibre. This is a generalisation
of the usual temporal gauge $\del_0 A_0=0$ of a gauge field on a circle, and
allows one to partially push the calculation down a dimension,
to the base of the Seifert fibration. And now further simplification
comes from the approach \cite{BT-CS, BT-Trieste-1993,
BT-S1,BT-Seifert,BKNT,BT-CCS} to evaluating the path integral through
Abelianisation.  
For Chern-Simons theory with gauge group $G$ the
end result is a finite dimensional integral over the Cartan subalgebra
$\lt$ of the Lie algebra $\lg$ of $G$.  
One of the characteristic 
features of this approach is that the reduction to
a finite-dimensional integral (over a linear space) 
completely bypasses the need to 
define (and integrate over) the moduli space of flat connections.
Nevertheless, this approach can be shown to reproduce the results of 
perturbative or localisation calculations (when available), which in principle 
require an exact evaluation of the latter.

Returning to $BF$ theory, the main technical result of this paper is to show
that this calculational method can be extended to $BF$ theory with its 
non-compact gauge group $TG$. The result is 
a finite-dimensional integral over the space $\lt
\oplus \lt$, which plays the role of the Cartan subalgebra of the
Lie algebra of $TG$. This once again bypasses the need to integrate
over the moduli space of flat connections. 
The finite-dimensional integrals that we find 
are similar to those for Chern-Simons theory, but they are more singular
than their Chern-Simons counterparts,
having poles on the integration contour an integral over
$\lambda$ which is a distribution, and hence require a
definition. For example, the partition function for $SU(2)$ and $M$ a
$\mathbb{Z}$ Homology Sphere devolves to
\be
Z_{BF}[M,SU(2)] \simeq \sum_{n_{0}} \int_{\mathbb{R} \times \mathbb{R}}
\tau_{M}\left(\phi\right)\, . \, \exp{\left(i\lambda \left( 
         \phi/P+2\pi i n_{0}
    \right) \right)} \label{BF-Seifert-intro}
\ee
(see the body of the paper for all the definitions). The important
point to note here is that the Ray-Singer Torsion $
\tau_{M}\left(\phi\right)$ has poles at $\phi =m \pi$ $m \neq sP$ and
zero when $m$ 
is proportional to $P$.  The $\lambda$ integral sets $\phi = Pn_{0}$
which formally avoids the poles. Still we can only conclude what the
integral should be after it is properly regularised (defined).

In the case of Chern-Simons theory the integrals that we find there
also have poles, however, there we could argue that the contributions
at the poles should not be included (lying on the walls of the Weyl
chamber) and we introduced a mass regulator that essentially allowed
us to avoid them. Unfortunately, in the present setting of $BF$
theory, we do not have a guiding principle and so we offer 3 different
possible definitions of $BF$ theory and discuss their advantages
as well as their shortcomings:

\begin{enumerate}
\item Direct Definition via Residues

The first, and the one we spend most time on (section
\ref{section-4}), is through the direct 
definition of the finite-dimensional integral \eqref{BF-Seifert-intro}.
In order to get a handle on the poles of the Ray-Singer Torsion on
$\lt \times \lt$, one defines the theory in such a way that that
it is given by the residue of all the possible poles including delta
function contributions. We find that, rather remarkably, this does
reproduce the expected contributions of  \textit{reducible} flat
connections which we can follow by making use of the Massive
Ray-Singer Torsion. In particular we show in detail how this gives
the expected results, as a sum over contributions from the isolated
non-trivial flat connections, for Lens spaces. Indeed, this definition
appears to capture the essence of the contribution of reducible
connections in general.
However, this approach does not reproduce the expected result
\eqref{IZ-plus} for connections that are isolated and irreducible. We
note that, suprisingly, some of the poles on the contour correspond to irreducible
flat connections, while others correspond to `complex' flat
connections. We offer a brief attempt at an explanation (related to gauge fixing)
for why the irreducible connections arise in this way.
\end{enumerate}

In view of this, we approach the problem from a different point of view 
in section \ref{section-5}, based on the fact that the gauge group
$TG$ can be regarded as a contraction of either the compact group
$G\times G$ or the complex group $G_\CC$. 
Correspondingly, the other 2 definitions that we consider 
are to regard $BF$ theory as arising either in
the large $k$ limit of $G \times G$ Chern-Simons theory at levels $k$
and $-k$, or in certain limits of the coupling $t=k+is$ of 
$G_{\mathbb{C}}$ Chern-Simons theory. This regularises the theory to some 
extent and also side-steps the gauge-fixing issue alluded to above.

\begin{enumerate}
\addtocounter{enumi}{1}
\item Definition via a large $k$ Limit of $G\times G$ Compact Chern-Simons Theory

The advantage of using the compact $G \times G $ theory is that the
Hilbert space of states is finite dimensional and only goes over
to the infinite dimensional Hilbert space of states of $BF$ theory
in the limit. This therefore acts as a natural regulator. A
disadvantage is that certain conditions must be met by the connections
so that the correspondence that we need exists. We offer examples
of Brieskorn $\mathbb{Z}$ Homology Spheres (Seifert manifolds with
3 exceptional fibres) \cite{Freed-Gompf} where the conditions are
met and one obtains the partition function of $BF$ theory
in the expected form \eqref{IZ-plus}.  These conditions are not met by all manifolds,
however. For example, the equality between $BF$ theory and the large $k$
limit of the $G \times G $ theory fails for certain Lens spaces.
As the Lens spaces are $\mathbb{Q}$ Homology Spheres one would like
to conjecture that the correspondence holds for isolated flat
connections of $\mathbb{Z}$ Homology Spheres in general.

\item Definition via a large $k$ Limit of $G_\CC$ Complex Chern-Simons Theory

As for $G_{\mathbb{C}}$ Chern-Simons theory
\cite{Witten-CCS,Gukov, DGLZ} with complex
level $k+is$, there are many formal correspondences with $BF$
theory. The most obvious is to set $k=0$
and let $s\rightarrow \infty$, which leads directly to the $BF$
action (\ref{BF}) and its non-compact symmetries (\ref{BF-symm}).
Given the difficulties we face with this action, this is not the
limit that we consider. Rather, we take $s=0$ and $k\rightarrow
\infty$. An advantage of this approach is that the finite
dimensional integrals that arise are slight variants of those
discussed by Lawrence and Rozansky \cite{LR} for $SU(2)$ and for
general $G$ by Mari\~{n}o \cite{Marino} in the context of $G$
Chern-Simons theory.  In principle then the same strategies that
apply there can be used here, though we leave that for the future.

To the extent that the perturbative large $k$ evaluation of
Chern-Simons theory reduces to a sum over contributions from flat
connections, this is then also true for the partition function of
$BF$ theory.
Overall, the definition of $BF$ theory via $G_\CC$
Chern-Simons theory appears to be the most complete (however, one
needs to ensure that one does not count complex flat connections
that cannot be conjugated into flat $G$ connections).

\end{enumerate}

One approach not followed here but which is certainly of interest is
resurgence \cite{Witten-Bonn}. Gukov,
Mari\~{n}o and Putrov  \cite{GMP-resurge} show that for $SU(2)$ one can start with
the Abelian contribution to the Chern-Simons partition function in the
large $k$ limit and Borel resum in order to obtain contributions from
non-Abelian flat connections. Once in the complex space of connections
there is the possibility that $SL(2, \mathbb{C})$ connections which are
not (conjugate to) $SU(2)$ flat connections will contribute. They show
that miraculously this does not happen. The importance of this for us
is that, as we will see, we appear to obtain contributions from
connections in our first approach that would, from this point of view
be considered to be `complex'. 

This paper is organised as follows. Section \ref{section-2} follows
Witten's \cite{WBF} approach to obtaining the partition function for
isolated irreducible flat connections. There we also consider what
formally happens when the connections are either not isolated or are
reducible. In these cases we can profitably make use of the 
massive Ray-Singer Torsion introduced in \cite{BKT-MRST}. This allows
us to express our expectations for the form of the partition function
in concrete terms.

In Section \ref{section-3} we formulate $BF$ theory
on Seifert 3-manifolds which are either $\mathbb{Z}$ or $\mathbb{Q}$
homology spheres. Particular attention is paid to issues with gauge
fixing of $\phi$ and $\lambda$, the components of $A$ and $B$ along
the fibre respectively. We will shows that 
$\phi$ and (with an assumption about $\phi$) 
$\la$ can be chosen to be constant along the fibre (temporal gauge), and to 
simultaneously take values in a fixed Cartan subalgebra $\lt \oplus \lt$ of 
$TG$ (Abelianisation in $TG$). We then give a brief outline of how to pass from the
functional integral to a finite dimensional integral over $\lt \oplus \lt$. 

As already mentioned, the integral in question has poles
on the integration contour and so requires a definition. In section \ref{section-4}
we thus give our first
definition of $BF$ theory, as a residue of a particular function
related to the integrand of the finite dimensional integral, and analyse its consequences.
Given the shortcomings of that definition, the alternative 
Definitions 2 and 3 described above are explored in section \ref{section-5}.

Certain technical details are relegated to an Appendix (\ref{appa} for information 
about the group $TG$ and its Lie algebra, and \ref{appb} for details about a certain
useful discrete symmetry of complex Chern-Simons theory on Seifert manifolds).

\section{$\boldsymbol{BF}$-Theory  on a 3-Manifold and Path Integrals}\label{section-2}

In this section we give a brief review of 3-dimensional $BF$
theory and Witten's formula for the path integral which holds when
there are no zero modes. This is followed by a discussion of a
generalisation which takes zero modes into account.

\subsection{$BF$ Theory as a $TG$ Chern-Simons Theory}

As recalled in the Introduction, 3-dimensional $BF$ theory 
on a 3-manifold $M$ is defined by the action \eqref{BF}
\be
I_{BF}= \int_{M} \Tr{\left( B \wedge F_{A}\right)} 
\label{2BF}
\ee
with $A$ a connection on a $G$-bundle over $M$ and $B
\in \Omega^1(M,\lg)$. The gauge symmetries of this action are the 
usual $G$ gauge symmetry acting on $A$ and $B$, 
as well as a shift gauge symmetry for the field $B$, 
\be
(A,B) \rightarrow (g^{-1}Ag + g^{-1}dg, \,
      g^{-1}\left( B 
  + d_{A} \Lambda \right) g) \label{BF-symm}
\ee
This shift gauge symmetry is non-compact, parametrised by $\Lambda\in \Omega^0(M,\lg)$, 
and thus, even for $G$ a compact Lie group, 3-dimensional $BF$ theory is 
an example of a gauge theory with a non-compact gauge group (and this is the main 
reason we are interested in this theory here).
In fact, $BF$-theory is not just some non-compact analogue of
Chern-Simons theory: in a precise sense it is a Chern-Simons theory
for the non-compact gauge group $TG$, the tangent bundle group of
$G$. As it will frequently be useful to have this perspective in
the back of one's mind in the following, here (and in more detail
in Appendix \ref{appa}) we quickly review the
relevant facts regarding the group $TG$, its Lie algebra $\ltg$,
and Chern-Simons theory based on it.

Thus, consider a compact semi-simple Lie group $G$ (throughout we will also assume
that $G$ is connected and simply-connected) with Lie algebra $\lg$, 
a basis of generators $j_a$ and commutation relations $[j_a,j_b]=f_{ab}^c j_c$. 
Then the Lie algebra $\ltg$ of the Lie group $TG$ has generators
$(j_a,p_a)$ and commutation relations 
\eqref{tga}
\be
[j_a,j_b] = f_{ab}^c j_c\quad,\quad [j_a,p_b] = [p_a,j_b] = f_{ab}^c p_c\quad,\quad [p_a,p_b]=0\;\;.
\ee
For the considerations of section \ref{section-5} it will be useful to keep in mind, that this algebra
can be obtained as a contraction of the Lie algebra of $G\times G$ or $G_{\CC}$ (see \eqref{tga2}).

Given an invariant non-degenerate scalar product $g_{ab} = \Tr j_a j_b$ on $\lg$, an invariant 
and non-degenerate scalar product on $\ltg$ is \eqref{second}
\be
\ll j_a,j_b\gg = \ll p_a,p_b\gg = 0 \quad,\quad \ll j_a,p_b\gg = g_{ab}\;\;.
\ee
A $TG$-gauge field can be expanded as 
\be
C = A^a j_a + B^a p_a \;\;,
\ee
with field strength
\be
F_{C} = dC + \trac{1}{2}[C,C] = F_{A}^a j_a +
d_{A}B^a p_a \;\;.
\ee
The Chern-Simons action for $C$ with respect to the above scalar product is
(with a for present purposes convenient choice of normalisation, and with an integration by
parts)
\be
I_{CS} \equiv \trac{1}{2}\int_M 
\ll C,dC+ \tfrac{1}{3}[C,C]\gg 
\;= \int_M \Tr B\wedge F_A = I_{BF}
\ee
The equations of motion \eqref{eq-motion} 
of $BF$-theory are then evidently simply the flatness conditions
for the connection $C$, 
\be
F_{C}=0 \quad\LRa\quad 
F_{A}=0, \;\;\; d_{A}B=0\label{2eq-motion}
\ee
Moreover, the gauge symmetries \eqref{BF-symm} of $BF$-theory are precisely the $TG$ gauge 
symmetries of the $TG$-connection $C$.

\subsection{Path Integral for $BF$ Theory: Formal Considerations}

Evidently (and as recalled in the Introduction), 
formally, the path integral for the partition fuction of $BF$ theory is
\begin{align}
Z_{BF}[M,G] & = \int DADB\, \exp{\left(i\int_{M}
  \Tr{\left( B \wedge F_{A
              }\right)}\right)}\nonumber\\
  & \simeq \int DA \delta\left(F_{A} \right)
\end{align}
so that we may expand about flat connections to give a more complete
evaluation. 
Following Witten
\cite{WBF}, we use the standard Faddeev-Popov covariant gauge-fixing 
procedure around these flat connections.
More details
about gauge fixing Chern-Simons theory with non-compact gauge groups
can be found in \cite{Bar-Natan-Witten}.
The
action and symmetries about a flat background connection $\omega_{(\alpha)}$ are
\be
I_{BF}= \int_{M} \Tr{\left( B \wedge F_{A+
      \omega_{(\alpha)}}\right)} \label{BF-again} 
\ee
\be
(A+\omega_{(\alpha)},B) \rightarrow
(g^{-1}(A+\omega_{(\alpha)} )g + g^{-1}dg, \,
      g^{-1}\left( B 
  + d_{A+ \omega_{(\alpha)}} \Lambda \right) g) \label{BF-symm-again}
\ee
while the covariant background field gauge
fixing conditions
\be
d_{{\omega_{(\alpha)}}}*A=0. \quad
d_{{\omega_{(\alpha)}}}*B=0
\ee
require us to make a choice of metric on $M$. Then the ghost and
gauge fixing action is 
\be
I_{GF}= \int_{M}\Tr{ \left(ud_{{\omega_{(\alpha)}}}*B +
    vd_{{\omega_{(\alpha)}}}*A + \overline{f}
    d_{{\omega_{(\alpha)}}} * d_{({\omega_{(\alpha)}}+A) }f + \overline{g}
   d_{{\omega_{(\alpha)}}} * d_{({\omega_{(\alpha)}}+A) }  g \right)} 
\ee

If the
space of flat connections is not made up of isolated points 
then one would need to integrate over them which one could do through
the introduction of collective coordinates. 
One should also take into account the possible
zero modes of both $B$ and of the ghosts. Those zero modes
form non-compact spaces $
\mathrm{H}^{1}_{\omega_{\alpha}}(M, \lg)$, and
$\mathrm{H}^{0}_{\omega_{\alpha}}(M, \lg)$ respectively.    
Taking the above caveats into account the path integral formally becomes
\begin{align}
Z_{BF}[M,G] &=\sum_{\alpha} \int DADB \,
\exp{\left(iI_{BF}+iI_{GF}  \right)} \nonumber\\
&=  \int\mathllap{\sum_{T\mathfrak{M}_{\alpha}}} \, \tau_{M}(\omega_{(
          \alpha)}) \label{TM-integral}
\end{align}
Here the sum and integral symbol over $T\frak{M}_{\alpha}$ is meant
to indicate that one sums over distinct components of the moduli
spaces of classical solutions and integrates over the moduli of
continuous families of solutions. These spaces include the zero
modes of $A$ and $B$ as well as of those of the
ghosts $f$ and $g$ (the multiplier field and anti-ghost zero modes
canonically cancel each other). 
The space $T\frak{M}_{\alpha}$ space can be very
singular and so (\ref{TM-integral}) as it stands is rather
symbolic in general. 

On the 3-manifolds of interest in this paper, the moduli space includes
acyclic flat connections, i.e.\ flat
connections which are isolated and irreducible, 
so that there are no zero modes at all
at
these solutions. The path integral around such a connection 
simplifies by using
the following scaling argument. About such an isolated and irreducible
connection send
\be
A \rightarrow t A, \quad B \rightarrow
t^{-1} B, \quad t \in \mathbb{R}_{+}
\ee
(with compensating scaling for the multiplier fields). This transformation
has trivial Jacobian, and the action remains well defined in the 
limit $t\ra 0$. Indeed, in this limit, 
\begin{align}
I_{BF} & \rightarrow \int_{M}
         Bd_{{\omega_{(\alpha)}}}A \nonumber\\
I_{GF}& \rightarrow \int_{M}  \Tr{ \left(ud_{{\omega_{(\alpha)}}}*B +
    vd_{{\omega_{(\alpha)}}}*A + \overline{f}
    * \Delta_{{\omega_{(\alpha)}}}f + \overline{g}*
   \Delta_{{\omega_{(\alpha)}}}  g \right)}
\end{align}
and the partition function around this connection is 
simply a standard path integral representation \cite{Schwarz,BT-BF}
of the Ray-Singer torsion
\cite{Ray-Singer} 
$\tau_M(\omega_{(\alpha)})$ of the flat connection $\omega_{(\alpha)}$, 
\be
\tau_{M}(\omega_{(
          \alpha)})= \left(\Det{ \Delta^{0}_{{
        \omega_{(
          \alpha)}}}} \right)^{3/2}. \left(\Det{
    \Delta^{1}_{{\omega_{(\alpha)}}}}\right)^{-1/2}  \label{Torsion}
\ee
(the superscripts
on the twisted Laplacians indicating the degrees of the forms on which
they act). Thus on such 3-manifolds, 
the partition function now becomes
\be
Z_{BF}[M,G] = \sum_{\mathrm{acyclic}} \tau_{M}(\omega_{(\alpha)})
 + \ldots 
\label{isolated-plus}
\ee
where the ellipses indicate the rest of the contributions to the path
integral. As we have already
explained, there may be zero modes to deal with in the extra terms
(\ref{isolated-plus}) which manifest
themselves as zeros of the determinants in (\ref{Torsion}) and which 
invalidate the simplifications that we were allowed to make
for the isolated irreducible connections.

One way to proceed is to adopt the prescription of Ray and Singer
\cite{Ray-Singer-Analytic} by first excising the zero modes and then adding
a correction term to ensure metric independence. This `extra' gauge
fixing of the zero modes may be achieved by a BRST procedure
\cite{BT-BF} the details of which have been eplained in some detail
in \cite{BKT-MRST}. The advantage of this method is that then the torsion
is a natural measure for the finite dimensional integral that appears
in (\ref{TM-integral}). While this defines the integrand, one is still
confronted with the problem of determining  and defining the space over
which this is to be integrated. Thus, 
while formally this appears to be a good definition, at a practical
level it seems somewhat intractable at the moment.

\subsection{Path Integral for $BF$ Theory and Massive Ray-Singer Torsion}

An alternative method for regularising such zero modes, 
and for keeping track of the ellipses in the formula \eqref{isolated-plus}
was advocated
in \cite{BKT-MRST}. The idea is to add a mass to the Laplacians that
appear in (\ref{Torsion}), thus lifting all the zero modes and side-stepping
(or at least initially bypassing) 
the problem of having to deal with them directly. As a first step, 
what this means is that
instead of using the twisted Laplacian $\Delta_{\omega}$ in the
definition of the Ray-Singer Torsion we instead use the
massive Laplacian $\Delta_{\omega} + \mu^{2}$ which now manifestly 
has a positive definite spectrum. 
The Massive Ray-Singer Torsion, on a 3-manifold,
for a flat connection $\omega$, is then defined to be
\be
\tau_{M}(\omega, \mu) = \left(\Det{(\Delta_{\omega}^{0} + \mu^{2})}\right)^{3/2}\,
. \, \left(\Det{(\Delta_{\omega}^{1} + \mu^{2})}\right)^{-1/2}
\ee
One can then define a regularised partition function by 
\be
Z_{BF}[M,G;\mu] = \int\mathllap{\sum} \, \tau_{M}(\omega_{(
          \alpha)}, \mu) \label{BF-implicit}
\ee
where we do not necessarily attempt to integrate over the tangent
space as those modes have been lifted. Here a caveat is required: in \cite{BKT-MRST} we showed
that for manifolds which are mapping tori it is possible to maintain
complete gauge invariance even with the introduction of a mass. However, a mass
term cannot be introduced in a gauge invariant way on a general
3-manifold. We accept that the gauge symmetry is broken and may be
reinstated if required by some renormalisation. 

The way in which one obtains the actual
Ray-Singer Torsion from the massive Ray-Singer torsion is
to take a limit
\be
\tau_{M}(\omega) \equiv \lim_{\mu \rightarrow 0} \mu^{-3\dim{\mathrm{H}^{0}_{\omega}}+
  \dim{\mathrm{H}^{1}_{\omega}}}\, \tau_{M}(\omega, \mu) 
\ee
In this way the $BF$ partition function will then take the form
\be
Z_{BF}[M,G,\mu] = \int\mathllap{\sum} \, \mu^{3\dim{\mathrm{H}^{0}_{\omega}}-
  \dim{\mathrm{H}^{1}_{\omega}}}\left( \tau_{M}(\omega) + \dots
\right) \label{int-sum-mu}
\ee
and we will need to specify which terms we are interested in. The
ellipses in this formula refer to terms higher order in $\mu$ than the
zero-th order Torsion (essentially constants and derivatives of the
Ray-Singer Torsion). 

At a formal level this is, perhaps, as far as one may go. It must
be said that this situation is not very satisfying and has calculational
drawbacks. The formula (\ref{BF-implicit}) is very implicit and
requires knowledge outside of the path integral in order to be used.
The flat connections need to be found, the cohomology groups about
the flat connections must be determined and a formula for the Massive
Ray-Singer Torsion must be given.

It is therefore of considerable interest to consider 3-manifolds
for which the BF partition function can also be calculated directly and
explicitly, and where the result can be compared with the formal 
expectations for the partition function outlined above.
One such class of manifolds is Seifert manifolds.

\subsection{Expectations for the Partition Function on Seifert
Manifolds}

The 
Ray-Singer torsion for Seifert 3-manifolds is known (through its equivalence to
Reidemeister Torsion \cite{Fried}). About an acyclic  flat
connection, i.e.\  with trivial twisted cohomology, it is
\be
\tau_{M}(\omega) =
\tau_{S^{1}}(\phi)^{2-N}\prod_{i=1}^{N}\tau_{S^{1}}(\phi/a_{i}) \label{RST-Seifert}
\ee
where $\phi$ is the component of the connection in the direction of
the fibre of the Seifert 3-manifold 
and $N$ (the number of orbifold points) and $a_i$ (the order of the isotropy 
group at the $i$'th orbifold point) are integers that are part of the defining 
data of a Seifert 3-manifold (see section \ref{subseifert}). 
We should also specify the
representation of the group $G$ for which we are evaluating the
Torsion; however, as throughout  we fix on the adjoint representation, 
we do not need to indicate it in the notation.

Including the contributions from the non-acyclic connections, 
one expects the partition function to take the form  \eqref{isolated-plus}
\be
Z_{BF}[M,G] \simeq \sum_{\text{acyclic}} \,
\tau_{S^{1}}(\phi)^{2-N}\prod_{i=1}^{N}\tau_{S^{1}}(\phi/a_{i}) + \dots\label{BF-anything}
\ee
and, as in the general situation 
above, one way to keep track of the ellipses in the above formula is 
to use the massive Ray-Singer torsion. In the case of Seifert manifolds
all we need is the massive Ray-Singer torsion on the circle, which is well
understood \cite{BKT-MRST}. Thus a
suitably regularised version of \eqref{BF-anything} is 
\be
Z_{BF}[M,G;\mu] = \sum \, \tau_{S^{1}}(\phi,
\mu)^{2-N}\prod_{i=1}^{N}\tau_{S^{1}}(\phi/a_{i}, \mu ) + \dots \label{BF-expand}
\ee
One can now use a mass expansion to identify the individual contributions to
the regularised partition function. 

The analysis that we have presented thus far, in the covariant gauges,
is an extension of that of Witten to those flat connections which are not
necessarily flat and isolated. One shortcoming with this approach, as
we have already explained, is that with a background field approach one
must, by independent means, find the flat connections and
determine the cohomology groups about them. One would prefer an approach that
evaluates the path integral directly and does not require the split
between the classical and quantum fields. The rest of the paper is
devoted to producing just such an approach. 

In particular, we will show in section  \ref{section-3} that the 
procedure developed in \cite{BT-CS, BT-Trieste-1993, BT-S1, BT-Seifert,BT-CCS, BKNT} to reduce the 
partition function of Chern-Simons theory for a compact gauge 
group $G$ to a finite-dimensional integral (over the Cartan subalgebra $\lt$ of $\lg$)
can be extended to the case at hand, namely $BF$ theory, 
or Chern-Simons theory with gauge group $TG$. 

The finite dimensional integrals have poles on the contour of
integration and so it then remains
to define and give a meaning to them. 
As we will see in section \ref{section-4}, 
a direct calculation of this finite-dimensional integral 
reproduces the expected form of the partition function 
only partially. In particular it captures the contributions of
reducible flat connections as expected but surprisingly does not give
the expected results for the isolated and irreducible flat connections.

This prompts us to investigate two other
possible definitions in Section \ref{section-5} which do lead to the expected 
form of the partition function discussed in this section.

\section{$\boldsymbol{BF}$-Theory on a Seifert Fibred 3-Manifold}\label{section-3}

In this section we specialise to Seifert fibred 3-manifolds. The extra
structure afforded by the $S^{1}$ principal bundle structure (over an 
orbifold base) allows
for convenient choices of gauge as well as for regularising the
Ray-Singer Torsion with the introduction of a mass term \cite{BKT-MRST}.
We are very brief with the background
material on Seifert manifolds as it has appeared before. We spend some
time on the gauge fixing as this is quite new for the $TG$ theories,
while the evaluation of the determinants is so close to that of the
determinants evaluated in
\cite{BT-CCS} that we borrow liberally from there.

\subsection{Seifert 3-Manifolds Briefly}
\label{subseifert}

From now on $M$ denotes a Seifert 3-manifold. Then
$M$ is a 3-manifold which is a circle $V$-bundle over the 2 dimensional
orbifold $\Sigma_{V}$ of genus $g$ with $N$ orbifold points. A general Seifert
manifold is written as $M[\deg{\mathcal{L}_{M}}, \, g,\, (a_{1}, \,
b_{1}), \dots , (a_{N}, \, b_{N})]$ where the $a_{i}$ are the
isotropies of the orbifold points, the $b_{i}$ are the weights of the
line $V$-bundle at the orbifold points and $\deg{\mathcal{L}_{M}}$ is the
degree of that line bundle. As the Seifert 3-manifold is the circle V-bundle of the line
V-bundle $\mathcal{L}_{M}$ it is also designated by $S(\mathcal{L}_{M})$.

Let $\mathcal{L}_{0}$ be a topological
line bundle at some smooth point on $\Sigma_{V}$ with degree 1 and $\mathcal{L}_{i}$
be the line V-bundles at the $i$-th orbifold points with degree
$1/a_{i}$ then
\be
\mathcal{L}_{M} = \mathcal{L}_{0}^{b_{0}} \otimes \mathcal{L}_{1}^{b_{1}}
\otimes \dots \otimes \mathcal{L}_{N}^{b_{N}}
\ee
and $M$ is the circle bundle of $\mathcal{L}_{M}$ \cite{Furuta-Steer} (we use normalised
Seifert data so that $1 \leq b_{i} < a_{i}$). 

The Seifert 3-manifold is smooth iff
$\mathrm{gcd}(a_{i}, \, b_{i}) =1 $ for each $i$. It is a
$\mathbb{Z}$ Homology Sphere ($\mathrm{H}_{1}(M , \, \mathbb{Z})=0$) iff the line
bundle $\mathcal{L}_{M}$ that defines it satisfies
\be
g=0, \;\;\; c_{1}(\mathcal{L}_{M})= b_{0} +
\sum_{i=1}^{N} \frac{b_{i}}{a_{i}} = \pm \frac{1}{a_{1} \dots
  a_{N}} \equiv \pm \frac{1}{P} \label{zhs} 
\ee
(here we have introduced the notation $P=a_{1} \dots a_{N}$). 
One consequence of these conditions is that $\mathrm{gcd}(a_{i}, \,
a_{j}) =1 $ for $i\neq j$. 
If one takes a tensor power of this line V-bundle,
$\mathcal{L}_{M_{d}}=\mathcal{L}_{M}^{\otimes d}$, then the
Seifert manifold $M_{d} = M/\ZZ_d$ is a
$\mathbb{Q}$ Homology Sphere ($\mathrm{H}_{1}(M_{d} , \, \mathbb{Q})=0$)  with
\be
g=0, \;\;\; c_{1}(\mathcal{L}_{M_{d}})= c_{1}(\mathcal{L}_{M}^{\otimes
  d})= \pm \frac{d}{a_{1} \dots 
  a_{N}} = \pm \frac{d}{P}\label{qhs} 
\ee
and
\be
|d| = |\mathrm{H}_{1}(M_{d}, \, \mathbb{Z})|
\ee

These are the 3-manifolds that we will exclusively concentrate on. The
reason for this choice of $M$ is two fold.

Firstly, if $M$ is neither a $\mathbb{Z}$ nor a $\mathbb{Q}$ homology sphere, then it
will necessarily have a non-zero dimensional moduli space of Abelian
flat connections which in turn means that there will be Abelian $B$
zero modes and hence non-compact directions to deal with.\footnote{Even if $M$
is a $\mathbb{Z}$ or $\mathbb{Q}$ homology sphere, there may be non-zero dimensional components of
the moduli
space of flat connections, but there are many $M$ for which the moduli
space of flat connections is made up of a finite number of points.} We do not want to
have to worry about such a situation here, as it is somewhat tangential to the other
issues that we wish to address. 

Far from all $\ZZ$ or $\QQ$ homology spheres are Seifert. 
The reason for choosing $M$ to be more specifically a Seifert $\mathbb{Z}$ or $\mathbb{Q}$ 
homology sphere is a more pragmatic one. Having $M$ a Seifert
manifold means that we have Fourier mode decomposition of all the
fields, so their components can be ultimately viewed as living on $\Sigma_{V}$,
and, as shown in  \cite{BT-CS, BT-Trieste-1993, BT-S1,BT-Seifert,BKNT,BT-CCS}, 
there are specific gauge choices available that
allow one to significantly simplify the evaluation of the partition function.

\subsection{$BF$ Action and Gauge Transformations on a Seifert Manifold}

As the Seifert fibred 3-manifold $M$ is a principal $U(1)$ bundle, it
also comes equipped with the fundamental vector field $\xi$ which generates the
$U(1)$ action. We also equip $M$ with a (nowhere vanishing) connection 1-form 
$\kappa$, i.e.\
\be
\iota_{\xi}\kappa = 1\quad,\quad L_\xi \kappa =  \iota_{\xi}d\kappa = 0\;\;,
\ee
and
\be
\int_{M} \kappa \wedge d\kappa = c_{1}\left(\mathcal{L}_{M} \right)\;\;.
\ee
Note that, acting on a form of any degree, one has 
\be
\iota_{\xi}\kappa + \kappa\iota_{\xi} =1
\ee
so that $\iota_{\xi}\kappa$ and $\kappa\iota_{\xi}$ are projection
operators onto the space of horizontal and vertical forms of fixed degree respectively.

Given a trivial $G$-bundle on a Seifert 3-manifold $M$, we correspondingly 
decompose the connection $A$ and 1-form $B$ of $BF$ theory
as
\be
A = A_{\Sigma} + \kappa \phi, \;\;\; B = B_{\Sigma} + \kappa \lambda
\ee
where
\be
A_{\Sigma} = \iota_{\xi} \kappa A, \;\;\; \phi = \iota_{\xi}
A, \;\;\; B_{\Sigma} =\iota_{\xi} \kappa B  ,\;\;\; \lambda
= \iota_{\xi}B
\ee
We may also decompose the exterior derivative as
\begin{equation}
  d  = \iota_{\xi} \kappa d + \kappa \iota_{\xi}d 
   = d^{\Sigma} + \kappa \iota_{\xi}d
\end{equation}
with twisted versions
\be
d_{A} = \iota_{\xi} \kappa d_{A} + \kappa
\iota_{\xi}d_{A} 
\ee
which acts on 0-forms as
\be
d_{A} \beta = d^{\Sigma}_{A} \beta + \kappa D_{\phi} \beta
\ee
In terms of this decomposition, the gauge transformations
\eqref{BF-symm} take the form, with $t$ a local fibre coordinate
\begin{align}
&  A_{\Sigma}\rightarrow g^{-1}d_{A}^{\Sigma}g, \;\;\; \phi \rightarrow
g^{-1}(\partial_{t} + \phi)  g \nonumber \\
&  B_{\Sigma} \rightarrow g^{-1}(B_{\Sigma} + d^{\Sigma}_{A}\Lambda)
g, \;\;\; \lambda \rightarrow g^{-1}\left(\lambda + D_{\phi}\Lambda
\right)g \label{gt}
\end{align}
and the action \eqref{2BF} becomes
\be
I_{BF}= \int_{M}\Tr{\left( B_{\Sigma} \wedge \kappa D_{\phi}A_{\Sigma}
    + \kappa _{\Sigma}B\wedge 
    d^{\Sigma} \phi + \lambda \kappa \wedge  F_{A_{\Sigma}} + \kappa\wedge
    d\kappa \phi \lambda\right)} \label{BF-Seifert}
\ee
with horizontal curvature 
\be
F_{A_{\Sigma}} = d^{\Sigma}A_{\Sigma} + A_{\Sigma}^{2}
\ee

\subsection{Gauge Fixing 1: Temporal Gauge for $A$ and $B$}\label{sec-fibre-independence} 

One of the great benefits of considering a Seifert manifold is that
there are very useful non-covariant gauges available. In particular, on a Seifert
manifold we can always choose the ``temporal gauge'' that the fibre-component $\phi$ of
$A$ is constant along the fibre, 
\be
 \iota_{\xi} d \phi =0 \;\;.
\ee
It turns out that (for generic $\phi$) one can also impose the analogouus condition on $\lambda$, 
\be
\iota_{\xi} d \lambda = 0 \;\;.
\ee
In local coordinates with $t$ a fibre coordinate, these conditions simply read
\be
\frac{\partial}{\partial t} \phi =0, \quad \mathrm{and} \quad \frac{\partial}{\partial
  t}\lambda =0 \label{gf-no-time}
\ee
The first of these can be achieved by having $\phi$ gauge equivalent
to some `time' independent field $\phi_{0}$; i.e.\  we have to solve
\be
 g(t)^{-1}(\del_t + \phi) g(t) = \phi_0 \quad
\LRa\quad 
\partial_{t}g(t) = g(t)\phi_{0}- \phi(t) g(t) 
\ee
This equation has the solution
\be
g(t) = P\exp{\left(\int_{t}^{0}\phi(s)ds\right)}\; \ex{t\phi_{0}} 
\ee
(where we have without loss of generality chosen $g(0)=1$). 
Periodicity $g(1)=g(0)$ of the gauge parameter now determines $\phi_0$ to be 
the average value of $\phi$ in the sense that 
\be
\ex{\phi_{0}} = P\exp{\left(\int_{0}^{1}\phi(s)ds\right)}
\ee
The second gauge condition, $\del_t\lambda = 0$, 
is perhaps not familiar and as there is a hidden
subtlety we will deal with it in some detail. Suppose we have already
implemented the first gauge choice (thus in the following $\phi=\phi_0$ is 
taken to satisfy $\del_t\phi=0$). To arrive at $\lambda$
constant along the fibre we need to solve
\be
\lambda + D_{\phi} \Lambda = \lambda_{0} \label{lambda-const}
\ee
for some $\lambda_{0}$ constant in $t$ to be determined. 
Writing 
\be
\del_t\left(\ex{t\phi} \Lambda\ex{-t\phi}\right) =\ex{t\phi} (D_\phi\Lambda) \ex{-t\phi} 
\ee
one sees that this equation has the solution. 
\be
h(t)\Lambda(t)h^{-1}(t) = \int_{0}^{t}\ex{s\phi}\,
(\lambda_{0}-\lambda(s))\, \ex{-s\phi}\, ds + \Lambda(0)
\ee
where
\be
h(t) = \ex{t\phi}, \quad h= \ex{\phi}
\ee
Periodicity in the gauge parameter $\Lambda(1)=\Lambda(0) $ gives us the equation
\be
(\Ad(h)-1) \Lambda(0) = \int_{0}^{1}\ex{s\phi}\,
(\lambda_{0}-\lambda(s))\, \ex{-s\phi}\, ds \label{gf-consistent}
\ee
Before going on we note that a shift of $\Lambda$ by a constant (in $t$) $\Lambda_0$
simply amounts to 
a shift in $\lambda_0$, 
\begin{equation}
\Lambda\ra\Lambda+\Lambda_0 \quad\Ra\quad \lambda_0 \ra \lambda_0 + [\phi,\Lambda_0]
\end{equation}
Without loss of generality we can therefore assume
that $\Lambda(0)=0$, so that the periodicity condition \eqref{gf-consistent} implies
\be
\int_{0}^{1}\ex{s\phi}\, (\lambda_{0}-\lambda(s))\, \ex{-s\phi}\, ds = 0\;\;.
\ee
The integral involving $\lambda_0$ can be done explicitly, 
with the result
\be
\frac{\ex{\ad(\phi)}-1}{\ad(\phi)} \lambda_{0} = \int^{1}_{0} \ex{s\, \ad(\phi)}\,  \lambda(s) ds\;\;.
\label{concon}
\ee
For \textit{generic} values of $\phi$, the operator on the left-hand side can be inverted to give 
\be
\lambda_{0} = \ad(\phi)(\ex{\ad(\phi)}-1)^{-1} \int^{1}_{0} \ex{s\,
  \ad(\phi)}\,  \lambda(s) ds
\ee
which tells us in which sense $\lambda_{0}$ can be regarded as  a $\phi$-weighted average of 
$\lambda$ over the fibre.

However, we see that this solution fails when $\phi$ is not suitably generic, so that the operator
$\exp\ad(\phi)-1$ has zero-modes. This should not come as a suprise, as this is precisely the 
condition that the operator $D_\phi = \del_t + \ad(\phi)$ has zero-modes, so that the 
gauge fixing condition \eqref{lambda-const} cannot be solved for all $\lambda(t)$. 
Indeed, for a zero
mode $\psi$ of $D_\phi$ one has (see also the discussion in section 2.3.4 of \cite{BKT-MRST})
\be
D_\phi\psi=0 \quad\Ra\quad \psi(t) = \ex{t\ad(\phi)}\psi(0) \;\;, \label{zmode-eq1}
\ee
and periodicity $\psi(1)=\psi(0)$ implies 
\be
\psi(1) = \psi(0) \quad\Ra\quad \left(\ex{\ad(\phi)}-1\right)\psi(0) =
0 \;\;.\label{zmode-eq2}
\ee
For $\phi$ taking values in the Cartan subalgebra $\lt$ of $\lg$, this is simply
the condition that $\phi$ has a component that is an element of the integral lattice $I(G)$ of $G$.

Normally this kind of constraint arising only for highly non-generic field configurations
would not be an issue as one is integrating over $\phi$ 
and the points where $\phi$ is integral have measure zero. 
Unfortunately, in $BF$ theory we get delta function support
onto particular values of $\phi$. In particular we will see when we
are considering $\mathbb{Z}$ Homology Spheres that these integral
points are the ones that are selected (cf.\ \eqref{phici}).

\subsection{Gauge Fixing 2: Abelianisation and Emergence of Non-Trivial 
Line Bundles}\label{ab-sec}

After having chosen the gauges (\ref{gf-no-time}), we still have gauge
transformations available that are constant along the fibre. For a compact
gauge group, one can locally use this residual gauge feedom to conjugate
the field $\phi$ into the Cartan subalgebra $\lt$ of $\lg$, i.e.\ one can 
impose the gauge condition 
\be
\phi^\lk=0 \quad\LRa\quad \phi = \phi^\lt \;\;,
\ee
where $\lk$ is the orthogonal complement of the Cartan subalgebra $\lt$ of $\lg$, 
\be
\lg = \lt \oplus \lk\;\;.
\ee
We will come back to the global issues involved in this choice of gauge below. 

However, first of all we need to address the question if we can do something
analogous in the case at hand, with gauge group $TG$. 
Since $TG$ is neither compact nor semi-simple, this is not a priori obvious. 
However, we show in Appendix \ref{subdiag} that one can use the adjoint action of 
$TG$ on its Lie algebra to conjugate any element into an element of $\lt \oplus \lt$, 
which plays the role of the Cartan subalgebra of the Lie algebra of $TG$. 
From one perspective, the reason we are able to do this is that 
$TG$ arises as the In\"{o}n\"{u}-Wigner contraction of a group where this
is certainly possible, namely $G\times G$.

Explicitly, this means that we can use the gauge transformation
\be
(\phi,\lambda) \mapsto (g, \Lambda)^{-1}.(\phi, \lambda) .(g, \Lambda) = (g^{-1}\phi g, g^{-1}(
\lambda + [\phi, \Lambda])g)
\ee
to (locally) impose simultaneously the Abelianising gauge conditions 
\be
\phi^{\lk} = 0 \quad,\quad  \lambda^{\lk}=0 
\quad\LRa\quad \phi=\phi^\lt\quad,\quad \lambda = \lambda^\lt\label{gf-k}
\ee
Now let us turn to the global issues regarding this gauge choice, which implies
that we will be dealing with an Abelian gauge theory on $\Sigma_V$. 
As has been described in detail in \cite{BT-Trieste-1993,BT-DIA}, 
there are topological obstructions to imposing the gauge choice on $\phi$ globally; 
in particular, conjugating $\phi$ to the Cartan subalgebra $\lt$ on $\Sigma_{V}$ is not
possible with smooth gauge transformations. However, if we insist on doing so
anyway, the price to be paid is that we must sum over all available
line bundles on $\Sigma_{V}$. The reason for the emergence of this sum has been
explained a number of times \cite{BT-DIA, BT-CS, BT-Trieste-1993, BT-S1, BT-Seifert, BKNT}. 

Since $TG$ is contractible to $G$ (and there are no topological issues involved in 
the shift transformation $\la \ra \lambda + [\phi, \Lambda]$ beyond those involving 
$\phi$ itself), there are no additional topological obstructions arising from 
diagonalisation in $TG$.

In practice there are two ways to introduce the non-trivial line bundles and the sum over them.
They both have merits and in any case are equivalent. The first makes contact with flat 
connections rather straightforwardly while the other is computationally easier.

We begin by describing the method using background fields described in more detail in \cite{BKNT}.
The available line V-bundles on $\Sigma_{V}$ are
\be
\mathcal{L}= \mathcal{L}_{0}^{\otimes n_{0}} \otimes
\mathcal{L}_{1}^{\otimes n_{1}} \otimes \dots \otimes 
\mathcal{L}_{N }^{\otimes n_{N}}\label{line-bundle-decomp-2}
\ee
and the basic idea is to allow for a non-trivial background connection on patches
about each divisor, with the divisors being the singular points and
with one divisor being a regular point of $\Sigma_{V}$. It was shown
in \cite{BKNT} that in the tubular neighbourhood $V_{(a_{i},b_{i})}$ of the $i$-th singular
point one could introduce a connection 1-form $\kappa_{i}$ which would
glue to a global one form $\kappa$ on $M$. In terms of these local forms and
the local surgery data
\be
a_{i}s_{i}-b_{i}r_{i}=-1, \quad \mathrm{with}\quad  (a_{i},b_{i})=1
\ee
the local orbifold data, the background
connection takes the simple form\footnote{A word on
  notation. Throughout we write connections in a basis of simple 
roots $\alpha_{i}$. For semi-simple simply laced Lie algebras we have
that the natural inner product 
$\langle \alpha_{i}, \alpha_{j}\rangle = 2\delta_{ij}$. The bold facing of integers
indicates that we mean $\mathbf{m}=m^{i}\alpha_{i}$ and for each $i$
we have
$m^{i} \in \mathbb{Z}$. We will write that $\mathbf{m} \in
I(G)$ where $I(G)$ is the integral lattice of $G$. As $G$ is simply
connected we have identified the root lattice with $I(G)$.}
\be
\mathcal{A}_{B} = 2\pi i \left( \mathbf{n}_{0} \kappa_{0} +
\sum_{i=1}^{N}\mathbf{n}_{i}  r_{i} 
\kappa_{i}\right)
\ee
where the $\mathbf{n}$ are Hermitian and lie in $\lt$.
With an abuse of notation, we may write the background as
\be
\kappa \, \phi_{B} = 2\pi i\left(\mathbf{n}_{0} \kappa_{0} +
\sum_{i=1}^{N}\mathbf{n}_{i} r_{i}\kappa_{i}\label{phi-back}
\right)
\ee
Each of the local forms $\kappa_{i}$ extends outside of the
tubular neighbourhood to $d\theta$ on the fibre. One also has that
\be
\int_{V_{(a_{i},b_{i})}} \kappa_{i}d\kappa_{i}= \frac{b_{i}}{a_{i}}
\ee
with the regular point having weight $a_{0}=1$. Glued together the
$\kappa_{i}$ form a global, smooth 1-form $\kappa = 
\kappa_{0}\cup \kappa_{1} \cup \dots \cup \kappa_{N}$ that
defines a principal bundle structure on $M$ with
\be
c_{1}(\mathcal{L}_{M}) = \int_{M}\kappa \wedge d\kappa = b_{0} +
\sum_{i=1}^{N} \frac{b_{i}}{a_{i}}
\ee
With the background field in place, one makes the substitution
\be
\kappa \phi \rightarrow \kappa \phi+ \kappa \phi_{B} \label{sub-phi}
\ee
everywhere in the path integral, with the second term on the right given by
(\ref{phi-back}).

Alternatively \cite{BT-CS, BT-Trieste-1993,
  BT-S1, BT-Seifert} one sets all the $\mathbf{n}_{i}$ $i\neq 0$ to zero and
$\kappa_{0} \rightarrow \kappa$ (and this is the natural procedure from the point of view of 
obstruction theory outlined in \cite{BT-DIA} and extended here to the
manifolds that we are considering). Otherwise one keeps the substitution
(\ref{sub-phi}) in the path integral. In Appendix \ref{appb}  we show that
there is enough symmetry in the theory to pass between these different formulations.

\subsection{Reduction of the Partition Function to a Finite-Dimensional Integral}

With all the preliminaries in place we can now outline how to perform
the path integral for the partition function.

Firstly the action (\ref{BF-Seifert}), taking into account the
non-trivial bundles that arise on Abelianisation, goes over to
\be
I_{BF}= \int_{M}\Tr{\left( B_{\Sigma} \wedge \kappa
    D_{\phi+\phi_{B}}A_{\Sigma} + \kappa B_{\Sigma} \wedge
    d^{\Sigma} \phi + \lambda \kappa \wedge  F_{A_{\Sigma}} + \kappa\wedge
    d\kappa (\phi+\phi_{B}) \lambda\right)} \label{BF-Seifert-2}
\ee
with $\kappa \phi_{B}$ and $\kappa d \kappa \phi_{B}$ suitably
understood. Indeed we note that the part of $B_\Sigma$ that is constant along the fibre
and takes values in the Cartan subalgebra $\lt$ only appears
through the term
\be
\int_{M}\Tr{\left( \kappa B_{\Sigma} \wedge d^{\Sigma} \phi \right)}\;\;.
\ee
Thus integrating out these modes of $B_{\Sigma}$ imposes the condition that $d^{\Sigma} \phi =0$.
Given the gauge conditions $\del_t\phi=0$, this implies that  $\phi$ is
constant on $M$. Likewise the component of $A_{\Sigma}$ constant along the
fibre and in the Cartan subalgebra only appears in
\be
\int_{M}\Tr{\left( \kappa A_{\Sigma} \wedge d^{\Sigma} \lambda \right)}
\ee
so integrating over that mode of $A_\Sigma$ fixes $d^{\Sigma} \lambda =0$ which, 
together with $\del_t\la =0$, tells us
that $\lambda$ is also constant on $M$. The other components of
$A^{\lt}_{\Sigma}$ and $B^{\lt}_{\Sigma}$ are paired with each other and lead to an
overall constant when integrated over.

We should also take care of any possible harmonic modes of the
components of $A$ and $B$ that we have integrated over. If there
any such harmonic modes of $B$, they would be non-compact directions
and would lead to divergences. In order to avoid this we fix on the
genus of $\Sigma_{V}$ to be zero, as in (\ref{zhs}) and (\ref{qhs})
thus ensuring that there are no $B^{\lt}_{\Sigma}$ zero modes to
contend with. The stronger choice that we have made of having $M$
be a $\mathbb{Z}$ or $\mathbb{Q}$ homology sphere is to ensure that
there are no moduli of Abelian connections on $M$ as then we would
necessarily have Abelian $B$ zero modes.

The form that we now have for the action is
\be
I_{BF}= \int_{M}\Tr{\left( B^{\lk}_{\Sigma} \wedge \kappa
    D_{\phi+\phi_{B}}A^{\lk}_{\Sigma} + \kappa \wedge  \lambda
    A^{\lk}_{\Sigma} A^{\lk}_{\Sigma}\right)} 
    + \Tr{\lambda \left( c_{1}(\mathcal{L}_{M}) \phi+ 2\pi i \widehat{\mathbf{q}}
    \right)} \label{BF-Seifert-2} 
  \ee
where
\be
\widehat{\mathbf{q}}=  \sum_{i=0}^{N}
        \frac{1}{a_{i}}\mathbf{n}_{i}
\ee

We are now  ready to integrate out the fields $A^{\lk}_{\Sigma}$ and $B^{\lk}_{\Sigma}$
that appear in (\ref{BF-Seifert-2}). There are also ghost terms of the
same type to take into account. The functional determinant that
we obtain and need to evaluate is
\be
\left. \Det{\left(\begin{array}{cc}
             \ad_{\lambda} & D_{\phi + \phi_{B}}\\
              - D_{\phi + \phi_{B}}         & 0
             \end{array}
  \right)}\right|_{\Omega^{1}_{H}(M, \lk) \oplus
             \Omega^{1}_{H}(M, \lk)}^{-1/2} \label{BF-dets}
\ee
There is a similar determinant from the ghost terms. As usual we will
expand in Fourier modes along the $S^{1}$ fibre and
we will regulate with a $\zeta$-function regularisation. What
one finds is, even upon regularisation, that the $\ad_{\lambda}$ part of the
determinant does not contribute. In this way what finally needs to be
evaluated is
\be
\left|\Det{\left(D_{\phi + \phi_{B}}\right)}\right|_{2\Omega^{0}_{H}(M,
  \lk)\ominus \Omega^{1}_{H}(M, \lk)} \label{dets}
\ee
For every Fourier mode $2\Omega^{0}_{H}(M,
  \lk)\ominus \Omega^{1}_{H}(M, \lk)$ is roughly the Euler characteristic of
  $\Sigma_{V}$. This, almost, cancellation of all the modes in the
  functional determinant can be used to evaluate it. However, the field
  $\phi_{B}$ is not constant on $\Sigma_{V}$ unlike $\phi$. This issue
  was circumvented in \cite{BKNT} by using a density form of the index
  theorem originally found in \cite{BT-CS}. The determinant in
  \cite{BKNT} is the square root of (\ref{dets}), so we may
  straightforwardly borrow the result from there. In this way we see
  that the absolute value (\ref{dets}) is
\be
\tau_{M}\left(\phi; \, \mathbf{n}_{i}\right) = \tau_{S^{1}}(\phi +
2\pi i \mathbf{n}_{0})^{2-2g
    -N}.\prod_{i=1}^{N} \tau_{S^{1}}((\phi + 2\pi i r_{i}
  \mathbf{n}_{i})/a_{i}) \label{rst}
\ee
where $\tau_{M}$ is the Ray-Singer torsion of $M$. The form of
$\tau_{M}$ shows that it is given by circles that is in terms of the Ray-Singer torsion
of $S^{1}$'s \cite{Fried}. For completeness we note that
\be
\tau_{S^{1}}(\varphi) = \mathrm{det}_{\lk}\left(I-\Ad{(\exp{\varphi})}
\right) \label{naive-RST}
\ee
so that $\mathbf{n}_{0}$ drops out of (\ref{rst}) since it has 
integral (lattice) entries.

We regularise the Ray-Singer Torsion by introducing a
mass \cite{BKT-MRST}. In the present context the mass is introduced into the
determinants (\ref{BF-dets}). This is not consistent with the $TG$
symmetry but does preserve the Abelian part of it. One exchanges (\ref{naive-RST}) with
\be
\tau_{S^{1}}(\varphi, \mu) = \mathrm{det}_{\lg}\left(I-\ex{-2\pi \mu}\Ad{(\exp{\varphi})}
\right) \label{MRST}
\ee
and it is understood that we use the massive form for the Ray-Singer
Torsion of the torsion on the circles on the right hand side of
(\ref{rst}). For the details of how this comes about see Example 2.5
in \cite{BKT-MRST}. We should note we have introduced extra factors of
$\mu$ (namely by the factor $(1-\ex{-2\pi \mu})^{\dim T}$) by passing to
(\ref{MRST}) and hence we have also changed the normalisation of the path integral

Putting everthing together, we see that we have now managed to reduce the 
path integral for the $BF$ partition function to a finite dimensional integral
over the (by now constant) fields $\phi$ and $\la$,
\be
Z_{BF}[M,G] = \sum_{\mathbf{n}_{0}} \dots
\sum_{\mathbf{n}_{N}}\int_{\lt \oplus \lt} \tau_{M}\left(\phi; \,
  \mathbf{n}_{i}\right)\, . \, \exp{\left(i\Tr{\lambda \left(
        c_{1}(\mathcal{L}_{M}) \phi+2\pi i \widehat{\mathbf{q}}
    \right)} \right)} \label{BF-Seifert-3} 
\ee
This is the integral that 
we will look at in detail in the the following section. 

\section{Definition 1: Direct Evaluation of the $\boldsymbol{BF}$
  (Path) Integral}\label{section-4}

As we saw in the last section we are able to perform the functional
integral representing the partition function $Z_{BF}$ to the point
that we are left with finite dimensional integrals. In this section we
analyse the remaining integrals and notice that there are issues with
them. Indeed this prompts us to search for a definition for
$Z_{BF}$. This definition does not agree with the perturbative result
of Witten \cite{WBF} in the case of isolated and irreducible
connections and we end the section by explaining where there
could be possible issues with one of the gauge conditions that we have
chosen. The formulae obtained capture those Abelian connections in
$\mathbb{Q}$ Homology Spheres and in these cases reproduce the
expected result (\ref{BF-expand}).

\subsection{Symmetries and other Properties of the Finite Dimensional Integrals}

At this point we have collected all the pieces and the $BF$
partition function, for the 3-manifolds under consideration, is
now a finite dimensional integral
\be
Z_{BF}[M,G] = \sum_{\mathbf{n}_{0}} \dots
\sum_{\mathbf{n}_{N}}\int_{\lt \oplus \lt} \tau_{M}\left(\phi; \,
  \mathbf{n}_{i}\right)\, . \, \exp{\left(i\Tr{\lambda \left(
       d \phi/P+2\pi i \widehat{\mathbf{q}}
    \right)} \right)} \label{BF-Seifert-3} 
\ee
where the integral over $\lt\oplus\lt$ is that over the constant Abelian 
``fields'' $\phi$ and $\lambda$ and we have used the fact that
$c_{1}(\mathcal{L}_{M})=d/P$ for the 3-manifolds we are considering as
given in (\ref{zhs}) and (\ref{qhs}) which in particular means that $g=0$.
There are no sums over the $\mathbf{n}_{i}$ $i\neq 0$ if we have set them
to zero, in which case we also have that $\widehat{\mathbf{q}}=
\mathbf{n}_{0}$. Note that as the gauge group $G$ is compact connected
and simply connected the $G$ bundles over $M$ are trivial and, on Abelianisation,
the line V bundles at the smooth point are `honest' line bundles so
that $\mathbf{n}_{0} \in \mathbb{Z}^{\mathbf{rk}}$. Consequently, in this case,
$\mathbf{n}_{0}$ only appears in the action in
(\ref{BF-Seifert-3}) (for either formulation of background).

In order to understand and ultimately evaluate the integral
\ref{BF-Seifert-3}, we note the following properties of the integrand:
\begin{enumerate}
\item both the action and the Ray-Singer torsion are invariant under the 
discrete symmetry 
\be
\phi \rightarrow \phi + 2\pi i P \mathbf{r}, \;\;\; \mathbf{n}_{0}
\rightarrow \mathbf{n}_{0} -d \mathbf{r}\;\;\; \mathrm{with}\; \; \mathbf{r}\in
\mathbb{Z}^{\mathbf{rk}(G)} , \;\; P = \prod_{i=1}^{N}a_{i}\label{phi-n-symm}
\ee
\item the Ray-Singer torsion with $N\geq 3$ has poles at
  $\alpha(\phi)= n \pi$ with $n \neq Pm$ (while for $N \leq 2$ there
  are no poles) \label{poles}
\item the Ray-Singer torsion has zeros at $\alpha(\phi)= m \pi P$
\item the Ray-Singer torsion is an even function of its arguments, in the sense that 
\be
\tau_{M}(\phi, \mathbf{n}_{i}) = \tau_{M}(-\phi, a_{i}-\mathbf{n}_{i})= \tau_{M}(-\phi, -\mathbf{n}_{i})
\ee
\end{enumerate}
Using the symmetry \eqref{phi-n-symm}, we may either `compactify' the field $\phi$ or limit the
range of $\mathbf{n}_{0}$. 
Either the way, the integral over $\lambda$ leads to a delta
function constraint on $\phi$, namely
\be
\phi=\phi_C \equiv -  \frac{2\pi i P}{d}
\widehat{\mathbf{q}} \label{class-phi}
\ee
and so it appears that only these values contribute to the path
integral. These values correspond to reducible connections as we have
seen in the past \cite{BT-CCS} and so one might conclude that the only
flat connections that contribute to the $BF$ partition function are
those that are Abelian. However, because of the presence of poles, 
this conclusion is a bit quick, and we will take a closer look at this
issue in section \ref{subsection-residue formula}.

Before moving on, it is worthwhile reminding ourselves that there
are restrictions on $M$ so the analysis we are performing holds.
As a counter-example, presume that $M$ is a product $\Sigma \times S^1$, 
or more generally a Seifert mapping torus,
i.e.\ with $c_{1}(\mathcal{L}_{M})=0$ (so definitely not one of the
manifolds that we are considering). Then there would be no symmetry
such as (\ref{phi-n-symm}) for the action and the integral over
$\lambda$ would not lead to (\ref{class-phi}).

\subsection{The $\phi_{\mathbf{C}}$ Contribution to $\mathbb{Z}$ Homology Spheres}

Here we specialise to the $\phi_{C}$ contribution to the
path integral for $M$ either an
$\mathbb{Z}$ or a $\mathbb{Q}$ Homology Sphere. We will use the symmetries
available not to limit the range of $\phi$ but rather to set
$\mathbf{n}_{0} \in I(G)/ d I(G)$ and we will keep the
$\mathbf{n}_{i}$ and so take the classical field
contribution to be
\be
\phi_{C} = -2\pi i \frac{P}{d}\sum_{i=0}^{N} \frac{\mathbf{n}_{i}}{a_{i}}
\ee
and any other choice leads to the same conclusions that we arrive at
presently. As we only want the $\phi_{C}$ contribution we
must ensure that $\tau_{M}$ has no poles and this is 
achieved by passing to the massive Ray-Singer torsion $\tau_{M}(\phi,
\mu)$ which we do.

If $M$ is a $\mathbb{Z}$ homology sphere then $d=1$ and the symmetry allows us to set
$\mathbf{n}_{0}=0$ so that
\be
\phi_{C} = -2\pi i P \sum_{i=1}^{N} \frac{\mathbf{n}_{i}}{a_{i}} \in I(G)
\label{phici}
\ee
Note that these are precisely the (potentially problematic) integral values of $\phi$ discussed
at the end of section \ref{sec-fibre-independence}. We will (have to) come back to this below.
A tiny bit of gymnastics\footnote{By (\ref{zhs})
$
Pb_{j}/a_{j} = 1 - a_{j}t_{j} , \; t_{j} \in \mathbb{Z}$
 and $
Pb_{j}r_{j}/a_{j} = r_{j} - r_{j}a_{j}t_{j} =
(1-a_{j}s_{j})P/a_{j} , \; t_{j} \in \mathbb{Z}$
whence $r_{j}-P/a_{j}= 0 \mod a_{j}\mathbb{Z}$. Consequently, $\phi_{C} +2 \pi i
\mathbf{n}_{j}r_{j} = -2\pi i P \sum_{i\neq j}
\mathbf{n}_{i}/a_{i} + 2\pi i \mathbf{n}_{j}.
(r_{j}- P/a_{j}) =0 \mod 2\pi i a_{j}I(G)$.} 
shows that
\be
 \phi_{C} + 2 \pi i
\mathbf{n}_{j} r_{j} = 0 \mod 2\pi i a_{j}I(G)
\ee

One immediately sees that the Ray-Singer Torsion (without a
mass) on the circle vanishes and the 
overall ratio in (\ref{rst}) is zero.
The poles and the zeros
of the finite dimensional integral arise as the zeros of the (inverse)
determinant of the $D_{\phi}$ operator. The poles are there when
  $D_{\phi}$ acts on 1-forms and the zeros arise when
acting on 0-forms at the special values (\ref{class-phi}) of $\phi$.

In order to understand the contributions we turn on the mass regulator, as in \eqref{MRST}.
Then we find
\be
\tau_{M}(\phi,\mu) =  \mathrm{det}_{\lg}\left(I - \ex{-2\pi\mu}\right)^{2-N}
  \prod_{i=1}^{N}\mathrm{\det}_{\lg}\left(I - \ex{-2\pi\mu/a_{i}} \right) 
\ee
The small mass limit gives us, as in (\ref{int-sum-mu}),
\be
\tau_{M}(\phi,\mu) = (2\pi \mu)^{2\dim{G}} . (a_{1}\dots a_{N})^{-\dim{G}}
+ \dots \label{red-torsion} 
\ee
The exponent of $\mu$ is understood as follows: the torsion of $M$ has
been pushed to that on $S^{1}$ and as on 
$S^{1}$ the twisted cohomology groups satisfy
$\mathrm{H}_{\omega}^{0}= \mathrm{H}_{\omega}^{1}$ the combination
$3\dim{\mathrm{H}_{\omega}^{0}}- \dim{\mathrm{H}_{\omega}^{1}} =
2\dim{\mathrm{H}_{\omega}^{0}}$ and
$\dim{\mathrm{H}_{\omega}^{0}}=\dim{G}$ for the trivial connection
which is the maximally reducible connection.

This result is quite far from the known Reidemeister Torsion for
regular elements of a Brieskorn Sphere $\Sigma(a_{1},a_{2},a_{3})$ \cite{Freed-Brieskorn},
\be
(a_{1}a_{2}a_{3})^{-1} \prod_{i=1}^{3}4 \sin^{2}{\left( \frac{n_{i}
      r_{i} \pi}{a_{i}}\right)}
\ee
in the case of $SU(2)$. The semi-classical analysis of Witten would
have suggested
\be
Z_{BF}[\Sigma(a_{1},a_{2},a_{3})\, SU(2)]  = (a_{1}a_{2}a_{3})^{-1}\prod_{i=1}^{3}
\sum_{n_{i}=1}^{a_{i}-1}4 \sin^{2}{\left( \frac{n_{i} 
      r_{i} \pi}{a_{i}}\right)} + \mathrm{reducible\, terms} \label{BF-su2}
\ee
 rather than the result that we obtained.

From (\ref{red-torsion}) we understand that
the $\phi_{C}$ contributions that we are looking at come from reducible
connections, for these manifolds the trivial connection as $G=SU(2)$, and so it
appears that we see only
the non acyclic terms in (\ref{BF-su2}). For example for $S^{3}$ the
only flat connection is the trivial connection and, as it has maximal
reducibility, one sees that a formula of the type
(\ref{red-torsion}) correctly captures this fact. Indeed, for any
(trivial) $G$ bundle over a $\mathbb{Z}$ Homology sphere the trivial
connection is flat and isolated and would give a contribution as
above. What are missing are the contributions of non-reducible or, for higher rank
groups, less reducible connections.

\subsection{The $\phi_{\mathbf{C}}$ Contribution to
  $\mathbb{Q}$ Homology Spheres}\label{subsection-Q-homology}

The zeros that were faced when $M$ is an $\mathbb{Z}$ Homology
Sphere are avoided to some extent when the manifold is a $\mathbb{Q}$ Homology
Sphere. The first advantage is that the points at which we wish to
evaluate the torsion are no longer necessarily at integer values but
rather at
\be
\phi_{C} = -2\pi i \frac{P}{d} \sum_{i=0}^{N}
\frac{\mathbf{n}_{i}}{a_{i}} \label{delta-function-points} 
\ee
Even if $\phi_{C}$ does at some points take integer values, the mass
regularised torsion remains well defined and that is what we will use.
We will give an example of this later on. Recall that here we fix the symmetry (\ref{phi-n-symm}) by insisting that
$\mathbf{n}_{0} \in I(G)/  d. I(G)$.

The $\phi_{C}$ contribution to the partition function is, with the
masses re-instated so that we may follow the reducibility of the connections,
\begin{align}
& \left. Z_{BF}[M,G, \mu]\right|_{\phi_{C}}   \nonumber\\
  & \;\;\; = \sum_{\mathbf{n}_{0}} \dots \sum_{\mathbf{n_{N}}}\tau_{S^{1}}(-2\pi i (P/d) \widehat{\mathbf{q}} +
2\pi i \mathbf{n}_{0} , \mu)^{2
    -N}.\prod_{i=1}^{N} \tau_{S^{1}}((-2\pi i (P/d) \widehat{\mathbf{q}} + 2\pi i r_{i}
  \mathbf{n}_{i})/a_{i}, \mu/a_{i}) \label{ZBF}
\end{align}
This result can be understood in terms of the homotopy representations
for flat connections on the Seifert $\mathbb{Q}$ Homology Sphere to which
we turn shortly. Notice that the contribution from these connections
is precisely of the form that was anticipated in
(\ref{BF-implicit}). However, for present purposes the most useful
form for us is where we have set the $\mathbf{n}_{i}$ for $i \geq 1$
to zero and restrict $\mathbf{n}_{0} \in I(G)/  d. I(G)$
\be
\left. Z_{BF}[M,G,\mu]\right|_{\phi_{C}} =
\sum_{\mathbf{n}_{0} \in I(G)/  d. I(G)}\tau_{S^{1}}(-2\pi i (P/d) \mathbf{n}_{0}, \mu)^{2
    -N}.\prod_{i=1}^{N} \tau_{S^{1}}(-2\pi i (P/d)\mathbf{n}_{0}
  /a_{i}, \mu/a_{i})\label{ZBF-QHS}
\ee

The classical examples of $\mathbb{Q}$ Homology Spheres are of course
the Lens spaces $L(p,q)$. The 
simplest Seifert presentation of the Lens space $L(p,q)$ is with
$N=1$, $d=p$ and $a=q^{*}$ where $qq^{*}=1 \mod p$, $p=b_{0}q^{*} +
b_{1}$. Substituting these values into (\ref{ZBF-QHS})
\be
Z_{BF}[L(p,q),G, \mu]  = \sum_{\mathbf{n} \in I(G)/  p. I(G)}
\tau_{S^{1}}(2\pi i
q^{*}\mathbf{n}/p + \mu). \tau_{S^{1}}(2\pi i  \mathbf{n}/p+ \mu/q^{*}) \label{L(p,q)}
\ee

To specialise to $G=SU(2)$ we note that the massive Ray-Singer Torsion (\ref{MRST})
takes the form
\be
\tau_{S^{1}}(\phi, \mu) = (1 - \ex{-2\pi \mu}). (1+ \ex{-4\pi \mu}
- \ex{-2\pi \mu} 2\cos{2\phi})
\ee
so that on specialising (\ref{L(p,q)}) to $G=SU(2)$ one obtains
\begin{align}
Z_{BF}[L(p,q), SU(2), \mu]  & = \left(\frac{\mu^{2}}{q^{*}}\right)^{3} +
                              \dots \nonumber\\
  & \; + \left(\frac{\mu^{2}}{q^{*}}\right)\sum_{n=1}^{p-1}
\sin^{2}{(2\pi i
q^{*}n/p )}. \sin^{2}{(2\pi i n/p)} + \dots \label{L(p,q)-SU2}
\end{align}
We note that happily (\ref{L(p,q)}) and (\ref{L(p,q)-SU2}) have the
form that we expected from the covariant analysis in Section
\ref{section-2} and in particular agree with the general form (\ref{int-sum-mu}).

The
first term in (\ref{L(p,q)-SU2}) corresponds to the trivial connection with
reducibility of order 3 while the second term is the sum over the
Abelian connections with reducibility of order 1 and the torsions agree with
the formulae obtained by Ray \cite{Ray} for Lens spaces. This should
also be compared with the Chern-Simons large $k$ limit formula 
(5.7) in \cite{Jeffrey} and we see that there is complete agreement
with the expression for the Ray-Singer Torsion of the Lens spaces for
the Abelian connections (the
only difference with Jeffrey's formula is that
we are summing over the Ray-Singer Torsion not its square root).

We note that $\phi_{C}$ in this case is the complete contribution to the
integral (\ref{BF-Seifert-3}) for, as noted in item \ref{poles} after
(\ref{phi-n-symm}), when $N=1$ there are no poles in the integrand and
the integral is well defined as is.

Now consider quotients of the Poincare sphere $\Sigma(2,3,5) =
M[-1, g=0, (2,1), (3,1), (5,1)]$. The fundamental group of
$\Sigma(2,3,5)$ is
$\pi_{1}(\Sigma(2,3,5))= I^{*}$ the binary icosahedral group of order
120 while the quotient manifold $\Sigma(2,3,5)_{d}=S(\mathcal{L}_{\Sigma(2,3,5)}^{\otimes
  d})$ has fundamental group $\mathbb{Z}_{d} \times I^{*}$
(Theorem 2 (v) page 112 in \cite{Orlik}). This means that $\Sigma(2,3,5)_{d}$ comes equipped with
the non-Abelian flat connections of $\Sigma(2,3,5)$ through the $I^{*}$ factor of
the fundamental group and with Abelian
connections through the $\mathbb{Z}_{d}$ factor of $\pi_{1}$.
In this case we have
\begin{align}
  & \left. Z_{BF}[\Sigma(2,3,5)_{d}, G,\mu]\right|_{\phi_{C}} \simeq \nonumber \\
  & \sum_{0 \leq \mathbf{n} \leq
  d-1} \tau_{S^{1}}(60\pi i \mathbf{n} /d
+\mu)^{-1}. \tau_{S^{1}}(30\pi i \mathbf{n}/d +\mu/2). \tau_{S^{1}}(20\pi i
    \mathbf{n}/d+ \mu/3) . \tau_{S^{1}}(12\pi i \mathbf{n}/d+
    \mu/5) \label{ZBF-Sigma(2,3,5)} 
\end{align}
Apart from having the standard form we note that we can expect, at
least in the case of $SU(2)$ that this formula correctly captures the
contributions of the reducible connections.

Indeed all the so called small Seifert manifolds with genus zero and
orientable base have finite fundamental groups with cyclic
factors. Hence quite generally the space of flat connections for these
manifolds naturally splits into the non-Abelian and Abelian parts. The
cyclic group is generated by the fibre generator $h$. For
large Seifert manifolds the fibre generator also generates the cyclic
group as a normal subgroup of the infinite order fundamental
group. Then $\mathbb{Z}_{d} \subset \pi_{1}(M_{d})$ and $\mathbb{Z}_{d}
= ( \pi_{1}/[\pi_{1}, \pi_{1}])(M_{d})$.

The interpretation of the formulae  at $\phi_{C}$  that we arrived
at above in terms of homotopy representations has been given in
\cite{BT-CCS}. At these reducible connections the Chern-Simons
invariant agrees with that given by Kirk and Klassen in \cite{Kirk-Klassen}, Auckly 
  \cite{Auckly} and Nishi \cite{Nishi} and is consistent with the
  identification of the connections being reducible and flat. Also the known
  relationship between the Ray-Singer Torsion for flat connections and 
  Reidemeister torsion corroborates the  identification of the
  contributions that we find.

\subsection{A Residue Formula for $SU(2)$}\label{subsection-residue formula}

As we saw in Section \ref{subsection-Q-homology} that Abelian
reducible connections are accounted 
for by $\phi_C$ we now take another and  closer look at the
finite-dimensional integral to see where the contributions from
non-Abelian  (ir)reducibles could come from.

The integrand of (\ref{BF-Seifert-3}) has poles and zeros and so we
must give meaning to the 
integral. A similar situation arises in Chern-Simons theory
\cite{BT-CS} where one explicitly removes the poles with the addition
of a mass term. There is a good reason for this in Chern-Simons theory
as, say on $\Sigma
\times S^{1}$, one is counting the finite number of states in the
theory. Here, if we use a mass regularisation then there is
the same effect, namely the poles do not contribute. However, this
then excludes important contributions to the $BF$ 
partition function. Also, it should be said, that for $BF$ theory one
does not have a finite dimensional Hilbert space of states and so
there is no `directive' to ensure that the poles should not be counted
in some way.

One could rotate the $\phi $ integral to say
$\exp{\left(-i\eps\right)} \times \lt$ which
would ensure that one avoids the poles altogether. However, one would need to
simultaneously require a different path of integration for $\lambda$
to ensure convergence of the integrals. Passing to the massive
Ray-Singer Torsion
is another way of avoiding the poles as the mass pushes the poles off
the real $\lt$ axis and into the complex plane $\mathbb{C}\times \lt$.

Short of a guiding principle we use the first symmetry in the Properties of
the Integrand to restrict the range of each component of the $\phi$
field to lie in $(-\pi P, \pi P)$. In this way we do not have to worry about
convergence issues with respect to integration over $\phi$. However,
there are still the poles to contend with and the integration over
$\lambda$.

In order to be concrete and to fix ideas we now focus on $G=SU(2)$. From the outset we
are tasked with having to make sense of 
\be
Z_{BF}[M,SU(2)] \sim \sum_{n \in \mathbb{Z}} \int_{\mathbb{R} \times \mathbb{R}}
\tau_{M}(\phi) \, . \exp{\left(i \lambda (\phi \frac{d}{P}- 2\pi n) \right)}
\ee
We firstly sum over $n$ to exchange $\lambda$ with the integers and
exchange the integral with the sum
\be
Z_{BF}[M,SU(2)] \sim  \int_{\mathbb{R}}
\tau_{M}(\phi) \, . \sum_{n \in \mathbb{Z}}\exp{\left(i n\phi \frac{d}{P} \right)}
\ee
As $\tau_{M}(\phi)$ is even in $\phi$ we have
\begin{align}
Z_{BF}[M,SU(2)] &\sim  \int_{-\pi P}^{\pi P}
\tau_{M}(\phi) \, . \frac{\left(1+ \exp{\left(i d\phi /P
      \right)}\right)}{\left(1 - \exp{\left(i d\phi /P
                  \right)}\right)} \nonumber\\
&\sim  \int_{-\pi P}^{\pi P}
\tau_{M}(\phi) \, . \frac{f(\phi)}{\left(1 - \exp{\left(i d\phi /P
                  \right)}\right)}  
\end{align}
and we have used the symmetries available to limit the range of
$\phi$ to lie in the range $(-\pi P, \pi P)$. Here the function $f$ is
\be
f(\phi) = \left(1+ \exp{\left(i d\phi /P
    \right)}\right)
\ee

There are now poles on the real axis coming both from the Ray-Singer
Torsion as well as from the result of the geometric sum. We push all
of these, in the complex plane, to lie above the real line (say by use
of an $i\epsilon$ prescription).

The contour that we choose in the complex plane in order to evaluate
the integral is as follows: travel along the real axis from $-\pi P$
to $\pi P$ then straight up along the imaginary axis to $\pi P+iR$
followed by an arc to $-\pi 
P+iR$ then straight down to $-\pi P$ and finally take the
$R\rightarrow \infty$ limit. For the integral over the arc to make sense
one regularises the integrand by multiplying by $\exp{(i \eps
  \phi^{2})}$ with $\eps >0$. The integrals over the other parts of
the contour are convergent as we will see below. The fact that the
prescription for the poles on 
the interval $(-\pi P, \pi P)$ are now such that they all lie inside
the contour (in the upper half plane) means 
that the contour integral is just given by the residue of all of those
poles.

Denote the contribution of each segment of the integral by the start
and end points as $I(x,y)$ so that we have, with $R \rightarrow \infty$
\be
I(-\pi P, \pi P) + I(\pi P, \pi P +iR) - I(-\pi P,-\pi P +iR ) = 2\pi i
\mathrm{Res} \left(\tau_{M}(\phi) \, . \frac{\left(1+ \exp{\left(i d\phi /P
      \right)}\right)}{\left(1 - \exp{\left(i d\phi /P
      \right)}\right)}  \right)
\ee
as the integral over the arc is zero. Had we preserved the symmetry
$\phi \rightarrow \phi + 2\pi P$ with our regulator then we would have
been able to declare that $I(\pi P, \pi P +iR) = I(-\pi P,-\pi P
+iR)$. However as $\phi$ becomes  large the integrand of $I(\pm\pi
P, \pm \pi P +iR)$ goes as
\be
\exp{\left(i\eps (i\phi\pm  \pi P)^{2} -2\phi (N-2-\sum_{i=1}^{N}1/a_{i}
)  \right)}\, + \dots 
\ee
and for $N \geq 3$ (but not for $\Sigma(2,3,5)$ the Poincare sphere)
there is convergence without the need for the Gaussian oscillatory behaviour so that
one may take the $\eps \rightarrow 0$ limit and the integrals over the
two vertical parts
of the contour cancel each other in the limit. In short, we have that
with this prescription for handling the poles that
\be
Z_{BF}[M,SU(2)] = 2\pi i
\mathrm{Res} \left(\tau_{M}(\phi) \, . \frac{\left(1+ \exp{\left(i d\phi /P
      \right)}\right)}{\left(1 - \exp{\left(i d\phi /P
      \right)}\right)}  \right) \label{SU(2)-res}
\ee

One would also
like to understand a little better what the residue formula
implies. To that end let
\begin{align}
\tau_{M}(\phi)& = \prod_{\alpha>0} \left(\sin{(\alpha(\phi))}\right)^{4-2N}
\, . \, \prod_{\alpha>0}
                \prod_{i=1}^{N}\left(\sin{(\alpha(\phi)/a_{i})}\right)^{2}\nonumber \\
  & = \prod_{\alpha>0} \left(\sin{(\alpha(\phi))}\right)^{4-2N}
\, . \, \widetilde{\tau}_{M}(\phi)
\end{align}

The residue formula in the case of $SU(2)$ takes the simple form
\begin{align}
  Z_{BF}[M,SU(2)] & =  Z_{BF}[M,SU(2)]|_{\phi_{C}} \nonumber\\
  & \,\; +
    \sum_{r=0}^{2N-5} \left(\begin{array}{c}
                              2N-5-r\\
                              r
                              \end{array}\right)\sum_{m= 1-P}^{P-1}\left(f(m\pi) . \,
              \frac{1}{1-\ex{im\pi d/P}}   \right)^{(2N-5-r)}
    \widetilde{\tau}_{M}(m\pi)^{(r)}
\end{align}
where the sum over $m$ indicates the poles of
$\left(\sin{(\alpha(\phi))}\right)^{4-2N}$. One sees that this is
essentially a sum over derivatives of the Reidemeister torsion
$\widetilde{\tau}_{M}$ (up to normalisation).

\subsection{$SU(2)$ $BF$ Theory on Brieskorn Spheres $\Sigma(a_{1},a_{2}, a_{3})$}

We now look concretely at $SU(2)$ $BF$ Theory on Brieskorn Spheres
$\Sigma(a_{1},a_{2}, a_{3})$, in order to see if the suggested
contour reproduces the expectations from perturbation theory as
derived by Witten \cite{WBF}. Even though the result for the path
integral is not exactly that predicted by Witten, we will see that this choice does have
contributions from irreducible flat connections which is quite pleasing in
its own right though, as we explain below, not completely unexpected.

With $N=3$ and $d=1$ the path integral becomes, apart from the
$\phi_{C}$ contributions of the previous sections
\begin{align}
  &Z_{BF}[\Sigma(a_{1},a_{2}, a_{3}), SU(2)] \nonumber \\
  & \simeq \sum_{m=1}^{P-1}
\left(\frac{1}{P}\csc^{2}{(m\pi/2P)} -4\cot{(m\pi/P)}
  \sum_{i=1}^{3}\frac{1}{a_{i}}\cot{(m\pi/a_{i})} 
\right) \widetilde{\tau}_{\Sigma(a_{1},a_{2}, a_{3})}(m \pi)\nonumber
  \\
  & \simeq \frac{1}{a_{1}a_{2}a_{3}}\sum_{m=1}^{P-1}h(m) \widetilde{\tau}_{\Sigma(a_{1},a_{2}, a_{3})}(m \pi)
\end{align}

To compare with (\ref{BF-su2}) we note that the integer $m$ can be written
in various ways in particular
\be
m = a_{1}\alpha_{1} + n_{1}r_{1}= a_{2}\alpha_{2} + n_{2}r_{2}=
a_{3}\alpha_{3} + n_{3}r_{3} , \quad \alpha_{i}, \, n_{i} \in \mathbb{Z}
\ee
to arrive at
\be
Z_{BF}[\Sigma(a_{1},a_{2}, a_{3}), SU(2)] \simeq
\frac{1}{a_{1}a_{2}a_{3}}\sum_{n_{i}} g(n_{i}) \, \prod_{i=1}^{3} 4\sin^{2}{\left(\frac{\pi n_{i}
r_{i}}{a_{i}}\right)} \label{BF-su2-n}
\ee
which would agree with (\ref{BF-su2}) if the function $g$ is $n_{i}$
independent.

As our first example (even though as we noted the derivation provided
fails) consider the Poincar\'{e} homology sphere $\Sigma(2,3,5)$. The
Ray-Singer-Torsion is the same for the values of $(n_{1},n_{2},n_{3})$
corresponding to $(1,1,r)$, $ (1,1,5-r)$ and $ (1,2,r)$. In this way
we have, apart from the trivial connection, the two possible
non-Abelian connections $(1,1,1)$ and $(1,1,2)$. By Proposition 2.8 of 
\cite{Fintushel-Stern}, this count is correct
for the flat connections on $\Sigma(2,3,5)$.

As our next example let us fix on $\Sigma(2,3,7)$ for which there are 2 irreducible
isolated flat connections. The Ray-Singer-Torsion is the same for
the values of $(n_{1},n_{2},n_{3})$ corresponding to $(1,1,r)$, $
(1,1,7-r)$ and $ (1,2,r)$. If we identify those contributions as
corresponding to the same isolated and irreducible connection then we
get 4 copies each of three different connections, say of $(1,1,1)$, $(1,1,2)$ and
$(1,1,3)$. In that case the sum in (\ref{BF-su2-n}) is over the 3
different connections each of which corresponds to 4 different
values of $m$ as tabulated below (the values of $g$ are approximate)
\be
\begin{array}{|c|c|c|}\hline
  m & (n_{1},n_{2},n_{3}) & g/P=\sum_{m}h(m)/P\\\hline
  1,13,29,41 & (1,1,1) & -34.45 \\\hline
  5,19, 23,37 & (1,1,2) & 6.96 \\\hline
  11,17,25, 31 & (1,1,3) & 3.35 \\ \hline
\end{array}
\ee
Hence
the partition function does not just give the sum of these 
connections with equal weight.

The situation is somewhat less clear than just indicated and
rather more puzzling as the third contribution $m=11,17,25, 31$ does
not correspond to a flat $SU(2)$ connection on $\Sigma(2,3,7)$. Rather
this connection is one identified in \cite{GMP-resurge} as a flat
$SL(2, \mathbb{C})$ connection that is conjugate to $SL(2,
\mathbb{R})$ and not to $SU(2)$ (cf.\ also \cite{Kitano-SL2R}). 

The holonomies of the isolated irreducible flat connections are
elements of the group and the holonomies can, independently, be conjugated into a
preferred maximal torus (even though they are not reducible). The
contribution to the Reidemeister Torsion of these holonomies is 
then determined by the integers $\mathbf{n}$ as described in
\cite{Freed-Brieskorn} equation 2.8 for Brieskorn spheres. To apply
this to an actual flat connection there are conditions on the integers arising
from the presentation of the fundamental group. What we have in our situation is
that all such integers arise as the poles of the Ray-Singer Torsion
and, unfortunately, our residue formula (\ref{SU(2)-res}) does not
restrict only to the ones
that correspond to honest flat $SU(2)$ connections.

\subsection{A Residue Formula For Higher Rank and Outlook}

Here we would like to briefly give a
formula in terms of residues for the partition function
(\ref{BF-Seifert-3}) to use as a definition for general $G$.

A possible definition, is 
\be
Z_{BF}[M,G] \simeq 2\pi i \mathrm{Res}\left(f(\phi) \tau_{M}(\phi)\, . \,
  \prod_{j=1}^{\dim \lt}
  \frac{1}{1-\ex{i\phi_{j}d/P}}   \right) \label{resZBF}
\ee
where $\phi = \phi_{j}\alpha^{j}$ and the function $f$ is such that
$f(2\pi \mathbf{n} P/d)=1$ for all $\mathbf{n}\in I(G)$,
\be
f(\phi) = \sum_{\mathbf{n} \in \Lambda} \ex{i n^{j}\phi_{j}
    d/P}\, f_{\mathbf{n}}
  \ee
The poles of $\prod_{j}(1-\exp{(i\phi_{j}d/P)})^{-1}$ are at $\phi_{C}$,
(\ref{class-phi}), so summing over its residues at these poles is
equivalent to performing the 
$\lambda$ integral. The
properties of $f(\phi)$ are so that at those 
poles $f$ is unity (whence the contribution is exactly the same as
performing the $\lambda$ integral). The function $f$ is obtained, as in the
$SU(2)$ case, by first performing the sum over $\mathbf{n}$ to restrict
the form of $\lambda$ and symmetry properties within the integral
\be
\sum_{\mathbf{n}} \int_{\lt} d\lambda \, \exp{\left(i\Tr{\lambda \left(
       d \phi/P+2\pi i \mathbf{n}
    \right)} \right)} \rightarrow \prod_{j=1}^{\dim \lt}
  \frac{f(\phi)}{1-\ex{i\phi_{j}d/P}}
\ee
It must also be remembered that the range of
$\phi$ is constrained.

The poles of $\tau_{M}$ may also contribute to the residue. If one takes
the attitude that one should use the massive Ray-Singer Torsion
directly then there are no extra poles and (\ref{resZBF}) agrees with
the naive evaluation of the path integral. Alternatively,
(\ref{resZBF}) allows one to take the poles of $\tau_{M}$ into account
after which one may reinstate the mass.

We have not shown that there is a contour that leads to (\ref{resZBF})
but one may reasonably hope that a generalisation of that used in the
$SU(2)$ case will be available.

Before ending this section we would like to suggest a possible apriori reason for why
the definition adopted here has shortcomings. We start by observing
that it is somewhat surprising that the possible isolated irreducible flat
connections are determined by poles in the Ray-Singer Torsion. This is
certainly counterintuitive as the poles of the Ray-Singer Torsion are
understood to arise when one has moduli, i.e. 1-form zero modes. A
possible reason for this is 
a shortcoming in the gauge-fixing choice that has been made. As already noted
in order to gauge fix $\lambda$ to be constant on the fibre we
require that $\phi $ be generic `enough'. To see that this may be the
cause of concern, imagine that one wishes, as we did in Section \ref{section-2},
to expand around an isolated irreducible flat connection. We do this but
we still insist on gauge fixing the quantum fields as we did in Section
\ref{section-3} so that $\lambda$ is constant along the fibre. Then we
would need to solve the equivalent of (\ref{lambda-const})
\be
\lambda + D_{\phi_{f}+\phi}\Lambda = \lambda_{0}
\ee
where $\phi_{f}$ is the fibre component of the background flat connection and
$\phi$ is the fibre component of the quantum field. If one thinks of
the quantum field as having a $\hbar$ in front of it then to lowest
order the equation to be solved is
\be
\lambda + D_{\phi_{f}}\Lambda \simeq \lambda_{0}
\ee
and the field $\phi_{f}$ is certainly not generic. The discussion
in Section \ref{section-3} tells us that precisely for these
$\phi_{f}$, as explained around (\ref{zmode-eq1}-\ref{zmode-eq2}), the
operator $D_{\phi_{f}}$ 
is not invertible and the 
gauge $\partial_{t} \lambda =0$ cannot be achieved.

Given this question about the gauge choice and the appearance of a
complex flat connection in the $SU(2)$ theory one may wonder if
resurgence is an approach that may demystify the situation. The $BF$
theory, as we have approached it here, arises as one possible limit of
$G_{\mathbb{C}}$ Chern-Simons theory (see the next section). In the
$SU(2)$ case the reducible connections are Abelian and one may hope
that the resurgence programme will yield the correct non-Abelian
connections as they do in \cite{GMP-resurge} with the correct
contribution to the partition function.

\section{Large $\boldsymbol{k}$ Asymptotics of $\boldsymbol{G \times G}$ and $\boldsymbol{G_{\mathbb{C}}}$ Chern-Simons Theory}\label{section-5}

Our aim in this section is to see if embedding $BF$ theory in a
`bigger' theory may act to give a suitable definition which can be
used in case $M$ is an $\mathbb{Z}$ Homology Sphere. To that end we
consider some naive large $k$ limits of Chern-Simons theory with the
compact gauge group $G \times G$ (with levels $k$
and $-k$ for the two factors) and the complexified group
$G_{\mathbb{C}}$. The `fattened' groups $G_{-k} \times G_{k}$ and
$G_{\mathbb{C}}$, as has already been explained (see Appendix
\ref{appa}), have been chosen as in the limits
to be discussed they both contract to $TG$.

The advantage of such an approach is that we avoid
the possible issues we had with the generic nature (or not) of $\phi$ as the
values of $\phi$ signalled out prior to taking the large $k$ limit
allow for the gauge where $\lambda$ is constant along the fibre.

In the case of $G_{-k} \times G_{k}$ we find that the theory does not
appear to `decompactify' enough in the large $k$ limit to capture all
the features of the $TG$ theory. However, in examples, we show that it
does actually pick up those flat connections which are both isolated
and irreducible correctly. The $G_{\mathbb{C}}$ Chern-Simons theory,
on the otherhand, has all the features of the $TG$ theory from the
outset. We show that the partition function has the formal properties
that would allow an application of the large $k$ analysis that have
worked in the case of the compact groups.

\subsection{Definition 2: $BF$ and the Large $k$ limit of $G_{k}\times$ $G_{-k}$ Chern-Simons Theory}

Here we consider the Chern-Simons action for two connections on
two copies of the same bundle for the
product gauge group $G \times G$ and for the following gymnastics we
require that the bundle in question is trivial. The action under
consideration is
\be
CS(A_{+}, A_{-})= \frac{1}{4\pi} \int_{M} \left[ \Tr(A_{+} \wedge dA_{+} +
  \frac{2}{3} A_{+} \wedge A_{+} \wedge A_{+})-\Tr(A_{-} \wedge dA_{-} +
  \frac{2}{3} A_{-} \wedge A_{-} \wedge A_{-}) \right]\label{AA-action}
\ee
The relative sign means that the two gauge theories have
opposite levels and we denote such a theory, with action $k\,
CS(A_{+}, A_{-})$ as a $G_{k}\times G_{-k}$ Chern-Simons theory. If we
denote the level $k$ Chern-Simons 
theory partition function for group $G$ by $Z_{CS}[M,G_{k}]$ then the partition function
for the $G_{k}\times$
$G_{-k}$ theory is
\be
Z_{CS}[M, G_{k}\times G_{-k}] = Z_{CS}[M,G_{k}]. Z_{CS}[M,G_{-k}]
\ee

Now to show how this is related to $BF$ theory make the substitutions
\be
A_{\pm} = A \pm \frac{\pi}{k} B , \quad A = \frac{1}{2}(A_{+}+A_{-}),
\quad B = \frac{k}{2\pi}(A_{+}-A_{-})\label{NC}
\ee
to arrive at
\be
CS(A_{+}, A_{-})= \int_{M}\Tr \left( B \wedge F_{A} +
  \frac{\pi^{2}}{k^{2}} B \wedge B \wedge B\right)
\ee
There may be a Jacobian $J$, depending on $k$ in passing to the new
variables in the path integral, so the relationship that we expect is
finally
\be
Z_{BF}[M,G] = \lim_{k \rightarrow \infty} J(k)\, Z_{CS}[M, G_{k}\times
G_{-k}]\label{ZBF-GkG-k}
\ee
where the limit as $k\rightarrow \infty$ formally ensures that the cubic term
in $B$ may be neglected. We allow $J$ to also soak up any other
factors of $k$ that may be present due to zero modes and so on.

This formal result says that the
$k\rightarrow\infty$ limit of a $G_{k}\times$ $G_{-k}$
Chern-Simons theory is equivalent to a pure BF theory. 
However, this statement certainly requires some elaboration as adding
and subtracting connections at will as in (\ref{NC}) is not 
usually an operation that makes sense in the theory of bundles. This is
mirrored in the gauge transformations that one obtains for $A$ and $B$
which are also not what one might call `standard'. To emphasise that
the limit in question is by no means obvious consider the 3-manifold
$\Sigma \times S^{1}$. Chern-Simons theory on such a manifold is
perfectly sensible and the large $k$ limit is well understood, the
partition function having leading term 
$k^{n/2}\mathrm{Vol}(\mathfrak{M}[\Sigma,G])$ the symplectic volume of
the moduli space of $G$ connections on $\Sigma$ where $n =
\dim{(\mathfrak{M}[\Sigma,G])} $
\cite{Witten-2d}. The large $k$ limit of the $G_{k}\times$ $G_{-k}$
partition function is then proportional to
$k^{n}\mathrm{Vol}(\mathfrak{M}[\Sigma,G])^{2} $ and  not what one would
expect from $BF$ theory, say
$k^{n}\mathrm{Vol}(T\mathfrak{M}[\Sigma,G])$ (even though this is ill defined).

So we need to better understand when the limit will indeed correspond
to $BF$ theory. At large $k$, the stationary phase
approximation tells us that the functional integral \cite{Witten-CS}
localises around the critical points of the action $CS(A)$,  
which in this case are precisely the flat connections. We then have,
for isolated flat connections, as k$\rightarrow$ $\infty$, 
\be
Z_{k} [M] \sim  \sum_{\alpha} \ex{2\pi ik CS(\alpha)} \,
e^{-\frac{i\pi }{2} I({\alpha})} \,
\sqrt{\tau_{M} ({\alpha})} \label{ZA}
\ee
where ${\alpha}$, $\tau_{M} (\alpha)$ and $I({\alpha})$ are the flat connections, 
the Ray-Singer torsion at the flat connection and the spectral flow to
the flat connection respectively.

The $G_{k}$ $\times$ $G_{-k}$ theory then has the asymptotic form
\be
Z_{k} [M] Z_{-k} [M] \sim
\sum_{\alpha, \beta} \ex{ik\left(CS({\alpha})-CS({\beta})\right)}
f_{\alpha \beta} \sqrt{\tau_{M} ({\alpha})} \sqrt{\tau_{M} ({\beta})} \label{ZkZ-k}
\ee
where 
\be
f_{\alpha \beta} = \exp{\left(-\frac{i\pi }{2} [I(A_{\alpha})-I(A_{\beta})]\right)}
\ee
One would expect therefore, that in the limit as $k\rightarrow \infty$,
that providing that flat connections $\alpha$ and $\beta$ having the
same Chern-Simons invariant implies they are the same flat connection,
\be
CS(\alpha) = CS(\beta) \, \mod
2\pi\mathbb{Z}\quad \implies \alpha=\beta \label{diff}
\ee
then the oscillations in
(\ref{ZkZ-k}) would ensure that we would indeed only need to sum  over
$\alpha=\beta$. In such a situation then one could reasonably expect
that (\ref{ZBF-GkG-k}) holds. If this condition does not hold then
 there may be `non-diagonal' contributions to the
sum (\ref{ZkZ-k}) and so the relationship with $BF$ theory becomes more tenuous.

Indeed for any connected component of non-isolated flat connections
all the flat connections in that component have the same Chern-Simons
invariant and fail our test (\ref{diff}). The flat connections on
$\Sigma \times S^{1}$ that we discussed previously are of this type
and this, to some extent, explains why they do not match the
expectations for $BF$ theory. The lesson here is that we must
concentrate on isolated flat connections. Fortunately for us
Fintushil and Stern \cite{Fintushel-Stern} have shown that the
Brieskorn spheres (the Seifert $\mathbb{Z}$ homology spheres with 3
exceptional fibres) have moduli spaces of flat $SU(2)$ connections
made up of a finite number of discrete points.

We shall now illustrate when (\ref{diff}) holds with a few examples in the
literature. Consider firstly the Poincar\'{e}
Homology Sphere, $M= \Sigma (2,3,5)$. The Poincar\'{e} $\mathbb{Z}$
homology sphere has three flat $SU(2)$
connections, one of which is the reducible trivial connection and
two which are non-Abelian. Freed and Gompf \cite{Freed-Gompf}
determine the large $k$ behaviour for the Chern-Simons partition
function and find
\be
Z_{k} [\Sigma(2,3,5)] \sim \sqrt{\frac{2}{5}} e^{-3\pi i/4}
\Big[\sin\Big( \frac{\pi}{5}\Big) e^{-\pi i (k+2)/60} +
\sin\Big(\frac{2\pi}{5}\Big) e^{-49\pi i (k+2)/60}\Big] 
\ee
Taking the modulus square, and presuming the limit of the norm
agrees with the norm of the limit, one gets
\be
\lim_{k\rightarrow \infty}Z_{k} [\Sigma(2,3,5)]Z_{-k} [\Sigma(2,3,5)] \sim \frac{2}{5}
\Big[\sin^{2}\Big( \frac{\pi}{5}\Big) +
\sin^{2}\Big(\frac{2\pi}{5}\Big)\Big] 
\ee
this corresponds to the sum of the Ray-Singer torsions of the
irreducible connections for $\Sigma
(2,3,5)$ up to a finite normalisation. Similarly \cite{Freed-Gompf}
provides us with the example of $\Sigma(2,3,17)$ and, with the same
caveat on exchanging limits and norms, we find that the norm square of
the Chern-Simons partition function is the sum of the Ray-Singer Torsion over the
6 irreducible flat $SU(2)$ connections.

There are manifolds, however, where you find two or more isolated flat
connections giving the same Chern-Simons invariant so that
(\ref{diff}) does not hold. Lens spaces yield examples of this
phenomena. As we have seen for Lens spaces $L(p,q)$ the
Ray-Singer Torsion is given by 
 \be
 \tau_{L(p,q)} = \frac{16}{p}\,  \sin^{2}\Big(\frac{2\pi n}{p}\Big)
 \quad \sin^{2}\Big(\frac{2\pi q^{*} n}{p}\Big) 
 \ee
 Here one integer is enough to describe the flat (Abelian) connections.

A typical example where two different connections have the same
Chern-Simons invariant is afforded by
the Lens space $L(12,5)$. For $L(12,5)$ we have $q^{*}=5$
 and one can see for example that $n=1$ and $n=5$ give the same
 Chern-Simons invariant (mod $2\pi$), namely 
 \be
 CS(n)= q^{*}  n^{2}/ p = \frac{5}{12} 
 \ee
 so we are not free to use the norm of the Chern-Simons theory in this
 case as a way to define the $BF$ theory.

The discussion of this section shows that this definition of $BF$
theory as a limit of the $G_{-k} \times G_{k}$ Chern-Simons theory
needs to be handled with care and we might have to know more
about the moduli space itself than we would care to. However, for the
part of the moduli space where the connections are isolated and
irreducible this appears to be an appropriate definition.

\subsection{Definition 3: $BF$ and the Large $k$ or $s$ Limit of $G_{\mathbb{C}}$ Chern-Simons Theory}
The final definition that we give involves the gauge group
$G_{\mathbb{C}}$. Note that the classical groups $TG$ and
$G_{\mathbb{C}}$ are diffeomorphic (as spaces). The
In\"{o}n\"{u}-Wigner contraction that establishes the Lie algebra
relationship is given in Appendix \ref{appa}. Furthermore, the
$G_{\mathbb{C}}$ connection is
\be
A_{\mathbb{C}}=A + iB
\ee
and as $G_{\mathbb{C}}$ contracts to the compact group $G$ the
connection $A$ may be considered a $G$ connection while $B$ is then a Lie
algebra $\lg$ valued one-form. This is a much clearer decomposition than the
mixing of objects that appeared in the $G_{-k}\times G_{k}$ theory of
the previous section.

The $G_{\mathbb{C}}$ Chern-Simons action is
\begin{align}
I(k,s) & =   \frac{k}{4\pi} \int_{M}  \Tr{\left( A\wedge d
    A + \frac{2}{3} A 
    \wedge A \wedge A - B \wedge
    d_{A}B \right)} \nonumber \\ 
&    \quad - \frac{s}{2\pi} \int_{M}  \Tr{\left( B \wedge
    F_{A} - 
    \frac{1}{3} B \wedge B \wedge
    B\right)} 
\end{align}
with $k\in \mathbb{Z}$ and we take $s \in \mathbb{R}$
\cite{Witten-CCS}. One can see that the complex Chern-Simons action has a number of
limits which formally lead to the $BF$ action. The first, and most straightforward way, to arrive at
the $BF$ theory is to set $k=0$ and consider the $s\rightarrow \infty$
limit where one sends $B\rightarrow B/s$ to formally arrive at
\be
I(k,s) \rightarrow - \frac{1}{2\pi}\int_{M}\Tr{\left( B \wedge
    F_{A} \right)} 
\ee
Alternatively one can consider the large $k$ limit, for finite $s$, and this time by
scaling $B \rightarrow B/\sqrt{k}$. One does not get the $BF$ action
directly, but rather
\be
I(k,s) \rightarrow \frac{k}{4\pi} \int_{M}  \Tr{\left( A\wedge d
    A + \frac{2}{3} A 
    \wedge A \wedge A\right)} - \frac{1}{4\pi} \int_{M}  \Tr{\left(B \wedge
    d_{A}B \right)} \label{CS-BF-ks}
\ee
About an isolated flat connection this would be the same as the large
$k$ limit of Chern-Simons theory except that the integral over the field $B$
has, formally the effect of replacing the square root of the
Ray-Singer torsion of the flat connection with the Ray-Singer torsion
itself and so would, in the limit, reproduce the partition function of
$BF$ theory with a Chern-Simons action contribution of the flat
connection. Both limits (and indeed any other limits of $k$ and $s$)
need to be carefully taken as in principle  one would most likely land
on the moduli space of flat $G_{\mathbb{C}}$ connections and not just the flat $G$
connections.

For finite $k$ and with $s=0$ Gukov and Pei \cite{Gukov-Pei} have
evaluated the partition function on $M=\Sigma \times S^{1}$. Even
though the Hilbert space is also infinite dimensional in this case, they ‘filter’
it by giving the $B$ field a mass and so are able to
extract meaningful results. Unlike the situation with the
$G_{-k}\times G_{k}$, where one gets essentially the square of the
compact case, here one sees that the results come from a non-compact
group and formally diverges as $k \rightarrow \infty$ as we would expect for the
$BF$ theory.

Returning to the case at hand note that the corresponding finite
dimensional integral to be performed in the case of complex
Chern-Simons after Abelianisation \cite{BT-CCS} has as its action
\begin{align}
 I(k,s)&= -\frac{k}{4\pi} \Tr(\phi^{2}
 -\lambda^{2})c_{1}\left(\mathcal{L}_{M}\right) + k \sum_{i=1}^{N}\frac{1}{a_{i}}
  \Tr (-i \phi 
\mathbf{n}_{i} + r_{i}\pi \mathbf{n}_{i}^{2}) + k \Tr(-i\phi
\mathbf{n}_{0}) \nonumber\\ 
&   + \frac{s}{2\pi} \Tr\Big( \lambda \left(\phi
 c_{1}\left(\mathcal{L}_{M}\right)
 +2\pi i\sum_{i=1}^{N}\frac{1}{a_{i}}
\mathbf{n}_{i} + 2\pi i  \mathbf{n}_{0}\right) \Big)\label{action-complex}
\end{align}
while the Ray-Singer Torsion goes over to the complex Ray-Singer
Torsion
\be
\tau_{M}(\phi) \rightarrow \sqrt{\tau_{M}(\phi+i\lambda)\tau_{M}(\phi-i\lambda)}
\ee
Formally the complex Ray-Singer Torsion under either scaling goes back
to the `real' Ray-Singer Torsion. The $k=0$ and $s\rightarrow \infty$
limit lands us on the finite dimensional action that we have been
using for $BF$ theory. Here we would like to investigate the
other limit.

The limit of interest then is large $k$ and for $\lambda\rightarrow
\lambda/\sqrt{k}$. As $\lambda$ now only appears in the action
quadratically one may integrate it out. In this limit the finite action goes over to
\be
 I(k,s)\rightarrow -\frac{k}{4\pi} \Tr(\phi^{2}
 ) c_{1}\left(\mathcal{L}_{M}\right) + k \sum_{i=1}^{N}\frac{1}{a_{i}}
  \Tr (-i \phi 
\mathbf{n}_{i} + r_{i}\pi \mathbf{n}_{i}^{2}) + k \Tr(-i\phi
\mathbf{n}_{0})
\ee
which is just the standard Chern-Simons action for compact group
$G$. The partition function, however, includes an extra factor of the
square root of the Ray-Singer Torsion (arising from the integral over
the field $B$ in (\ref{CS-BF-ks})) in the integral over $\lt$ so
this is not the partition function of compact Chern-Simons theory.

As the action agrees with that of $G$ Chern-Simons theory then all the
properties of the integrand (see Appendix \ref{appb} for the symmetry
properties) and hence the considerations that appear
in \cite{LR} and \cite{Marino} apply here too. This includes the 
choice of contour to take the large $k$ limit, the only difference
being that of
considering a slightly different function that takes into account that
we have the Ray-Singer Torsion not its square root in the
integrand. The resulting large $k$ asymptotics are, for
Chern-Simons theory, just as predicted by Witten in \cite{Witten-CS},
contributions around flat connections. Consequently, in the
limit that we are considering in the $G_{\mathbb{C}}$ theory, this
means that the asymptotic form will have contributions
from flat connections in $BF$ theory as anticipated in Section
\ref{section-2}. 

Specifically in the case of
$SU(2)$ the large $k$ limit is that of the flat
connection contributions on page 302 of \cite{LR} but with $F(y)$
there replaced with $F(y)^{2}$ in order to pass to the $BF$ theory (and one
should not include the framing as it ought not to arise in complex
Chern-Simons theory) while for a general compact
group one can refer to (4.17) in \cite{Marino} from which one can deduce the
large $k$ behaviour. At this point this particular definition of $BF$
theory is the most complete that we have.

If one would like to obtain more explicit formulae for the $BF$ theory
one could follow the explanation of \cite{LR} on how to arrive at
Rozansky's formula for the large 
$k$ expansion of $SU(2)$ Chern-Simons theory \cite{Rozansky-large-k}
which reproduces Witten's expansion. Of course this also encodes some
of the large $k$ structure of the $G_{\mathbb{C}}$ theory itself so that by
following different parts of the asymptotic expansion we would be able
to obtain parts of the perturabation theory. In particular we also
have in mind the contributions about the trivial connection. We leave these
issues, and the question of how and whether complex connections
contribute to the $BF$ limit for the future.

\section*{Acknowledgement}
M. Kakona thanks the External Activities Unit of the ICTP for a
 Ph.D. grant at EAIFR in the University of Rwanda and the HECAP group at the ICTP for
 supporting a visit to complete this work. He is also indebted to OFID
 and the STEP programme of the ICTP for funding and support in the
 initial stages of his studies. The group of M. Blau 
is supported by the NCCR SwissMAP (The Mathematics of Physics) of the
Swiss Science Foundation. 

\appendix

\section{The Tangent Bundle Group $TG$ and its Lie Algebra $\ltg$}
\label{appa}

\subsection{Basic Properties of $TG$ and $\ltg$}

let $G$ be a compact Lie group and (to not unnecessarily complicate things)
assume that $G$ is connected, simply-connected and semi-simple. 
Via a right-invariant
trivialisation of the tangent bundle, the group $TG$ can be identified with the semi-direct 
product of $G$ with its Lie algebra $\lg$, 
\be
\label{tgsdp}
TG \simeq G \ltimes\lg \equiv G \times_{\Ad}\lg\;\;,
\ee
with multiplication
\be
\label{tgm}
(g,v)(h,w) = (gh, v + \Ad_g w) \} \;\;,
\ee
where the adjoint action of $g\in G$ on $w\in \lg$ is $\Ad_g w = g w g^{-1}$. 
Correspondingly,  its Lie algebra has the form 
\be
\label{adtg}
\mathsf{Lie}(TG) \equiv \mathfrak{tg} \simeq \lg \oplus_{\ad}\lg_{Ab}
\ee
where $\lg_{Ab}$ is the Abelian Lie algebra based on the underlying vector space $\lg$, 
i.e.\ the commutator is
\be
\label{tgc}
[(x,v),(y,w)] = ([x,y],[x,w]+[v,y]) =([x,y],\ad_x w - \ad_y v)
\ee
and the (inverse) adjoint action of $TG$ on its Lie algebra is
\be
\label{TGAd}
(g,v)^{-1}(y,w)(g,v) = (\Ad_{g^{-1}}y,\Ad_{g^{-1}}(w + [y,v]))
\ee
Let $\ell_a$ be a (real) basis of the (real) Lie algebra $\lg$, with
\be
[\ell_a,\ell_b] = f_{ab}^c \ell_c \;\;.
\ee
Then
\be
j_a = (\ell_a,0)\quad,\quad p_a = (0,\ell_a) 
\ee
are a basis of $\mathfrak{tg}$, and the Lie algebra commutation relations take the form
\be
\label{tga}
[j_a,j_b] = f_{ab}^c j_c\quad,\quad [j_a,p_b] = [p_a,j_b] = f_{ab}^c p_c\quad,\quad [p_a,p_b]=0\;\;.
\ee

Here are two useful properties of $\ltg$:

\ben
\item The Lie algebra $\ltg$ of $TG$ can be obtained as a contraction of the 
Lie algebra of the group $G \times G$

\begin{itemize}
\item Consider the $G\times G$ Lie algebra 
\be
\lg_+ \oplus \lg_- = \lg \oplus\lg
\ee
with generators
\be
\label{gplusg}
[j_a^\pm,j_b^\pm] = f_{ab}^c j_c^\pm\quad,\quad [j_a^+,j_b^-]=0\;\;.
\ee
\item Perform the redefinition 
\be
\label{gplusg2}
j_a^\pm = \tfrac{1}{2}(j_a \pm p_a/\epsilon)
\quad\LRa\quad 
j_a = j_a^+ + j_a^-\quad,\quad p_a = \epsilon(j_a^+-j_a^-) 
\ee
\item Then the algebra $\lg \oplus\lg$ takes the form 
\be
\label{tga2}
[j_a,j_b] = f_{ab}^c j_c\quad,\quad [j_a,p_b] = [p_a,j_b] = f_{ab}^c p_c\quad,\quad [p_a,p_b]=\epsilon^2 
f_{ab}^c j_c
\;\;.
\ee
\item In the limit $\epsilon\ra 0$, this reduces to \eqref{tga}, the Lie algebra of $TG$.
\item If one sets $\epsilon = i\delta$, the same reasoning shows that $\ltg$ can be 
obtained as the contraction of the complexification $\lg_{\CC} = \lg \oplus i\lg$ of $\lg$. 

\end{itemize}

\item Existence of Invariant Scalar Products on $\ltg$

Let us consider the case where the Lie algebra $\lg$ is simple. In that
case, $\lg$ has a preferred (and up to a choice of scale unique) non-degenerate
and $\ad$-invariant metric scalar product, namely the Killing-Cartan form 
$\Tr \ad_x \ad_y$ (the trace in the adjoint representation). 
With respect to the basis $\ell_a$ of 
generators, this metric has the components
\be
\label{gab2}
g_{ab} \equiv <\ell_a,\ell_b> = \Tr\ad_{\ell_a}\ad_{\ell_b} = f^d_{ac}f^c_{bd}
\ee
Turning to the Lie algebra 
\be
\ltg\simeq \lg \oplus_{\ad} \lg_{Ab}\;\;,
\ee
as it is not semi-simple, its Killing-Cartan form
will be $\ltg$-invariant (by construction, i.e.\ by the Jacobi identity)
but degenerate. Indeed, in terms of the generators $(j_a,p_a)$ one has
\be
\Tr\ad_{j_a}\ad_{j_b} = 2 g_{ab} = 2 f^d_{ac}f^c_{bd}
\ee
(from the adjoint action of $\ad_{j_a}\ad_{j_b}$ 
on $j_c$ and on $p_c$), and
\be
\Tr\ad_{j_a}\ad_{p_b} = \Tr\ad_{p_a}\ad_{p_b} = 0
\ee
(because both  
$\ad_{j_a}\ad_{p_b}$ and $\ad_{p_a}\ad_{p_b}$ only act non-trivially on a $j_c$ and take it 
to a linear combination of the $p_c$).
This scalar product can therefore also be written as
\be
\label{first}
<j_a,j_b> = 2 g_{ab}\quad,\quad 
<j_a,p_b> = <p_a,p_b> = 0\;\;.
\ee

In addition to the (degenerate) Killing-Cartan form, $\ltg$ exceptionally also possesses 
a non-degenerate $\ad$-invariant scalar product given by 
\be
\label{second}
\ll j_a,j_b\gg = \ll p_a,p_b\gg = 0 \quad,\quad \ll j_a,p_b\gg = g_{ab}\;\;.
\ee
It is easily verified that this is indeed both invariant and non-degenerate.

The existence of this second invariant scalar product, or of the overall
two-parameter family of invariant scalar products, can be understood in
terms of the contraction of the Lie algebra $\lg \oplus \lg$ to $\ltg$
mentioned above. Indeed, 
starting with the non-degenerate Killing-Cartan metric on the Lie algebra 
$\lg \oplus \lg$ \eqref{gplusg} of $G \times G$, with coefficients $c^\pm$, 
\be
\label{gpmab}
<j_a^\pm,j_b^\pm> = c^\pm g_{ab}\quad,\quad <j_a^+,j_b^-> = 0 \;\;,
\ee
and performing the redefinition \eqref{gplusg2}, one finds
\be
<j_a,j_b> = (c^+ + c^-) g_{ab}\quad,\quad
<j_a,p_b> = \epsilon (c^+ - c^-) g_{ab}\quad,\quad
<p_a,p_b> = \epsilon^2 (c^+ + c^-) g_{ab}
\ee
Taking the contraction $\epsilon \ra 0$ with $c^+=c^- = 1$, one finds \eqref{first}, while
taking $c^+=-c^- = 1/2\epsilon$ one finds \eqref{second},
\be
\begin{aligned}
\label{cpcm}
c^+=c^- = 1 \quad&\Ra\quad <.,.> \\
c^+=-c^- = 1/2\epsilon \quad&\Ra\quad \ll .,.\gg
\end{aligned}
\ee

\een

\subsection{Diagonalisation / Abelianisation in $TG$ and $\mathfrak{tg}$} 
\label{subdiag}

Let $G$ be compact, $T_G$ be a maximal torus, $\mathfrak{t}_G$ its Lie algebra, 
a Cartan subalgebra. Then the following assertions are true:
\ben
\item
For any $h \in G$ one can find a $g\in G$ such that 
\be
h \in G \quad\Ra\quad \exists g\in G: \quad \Ad_gh = g h g^{-1} \in T_G\;\;.
\ee
\item
For any $y \in \lg$ one can find a $g \in G$ such that 
\be
y \in \lg \quad\Ra\quad \exists g \in G:\quad \Ad_g y = gyg^{-1}\in \lt_G\;\;.
\ee
\een
As $TG$ is not compact and not semi-simple, a priori the corresponding statements
do not necessarily hold (and would usually be far from true for a generic such group). 
Nevertheless, it turns out that the 
above statements carry over literally to the case of the group $TG$, provided that 
we replace the maximal torus $T_G$ of $G$ and the Cartan subalgebra $\lt_G$ of $\lg$ 
by
\begin{align}
T_{TG} &= T(T_G) \simeq T_G \times_{\Ad} \lt_G = T_G \times \lt_G\\
\lt_{TG} &= \mathsf{Lie}(T_G\times \lt_G) = \lt_G \oplus \lt_G
\end{align}
(since $\lt_G$ is already an Abelian Lie algebra, here it is not necessary to 
write $(\lt_G)_{Ab}$ in the second factor / summand). 
Indeed, we can now prove the following two statements:
\ben
\item Diagonalisation / Abelianisation in $TG$
\be
\forall (h,w)\in TG \simeq G \times_{\Ad}\lg  \quad\exists (g,v)\in TG:\quad 
(g,v)^{-1}(h,w)(g,v) = (t,\tau_2) \in T_G \times \lt_G 
\ee
\item Diagonalisation / Abelianisation in $\ltg$
\be
\forall (y,w)\in \mathfrak{tg}\simeq \lg \oplus_{\ad}\lg  \quad\exists (g,v)\in TG:\quad 
(g,v)^{-1}(y,w)(g,v) = (\tau_1,\tau_2) \in \lt_G \oplus \lt_G 
\ee
\een

Here are the proofs of these assertions:

\ben 
\item Diagonalisation / Abelianisation in $TG$

\begin{itemize}
\item 
The (inverse) Ad-action of $TG$ on itself is 
\be
\label{Adgroup}
(g,v)^{-1}(h,w)(g,v) = (\Ad_{g^{-1}}h,\Ad_{g^{-1}}(w + \Ad_{h}v-v))
\ee
In particular, for the action of $(g,0)$ one finds
\be
(g,0):\quad h \mapsto \Ad_{g^{-1}}h\quad,\quad w \mapsto \Ad_{g^{-1}}w
\ee
and for that of $(e,v)$ one has
\be
(e,v):\quad h \mapsto h\quad,\quad w \mapsto w + \Ad_h v - v\;\;.
\ee

\item First of all, we can and will choose $g\in G$ such that 
\be
h \mapsto g^{-1}hg=t \in T_G\;\;.
\ee
\item 
Next we define $S$ to be the stabiliser of $t$ under the adjoint action of $G$, 
\be
S:=\mathsf{Stab}_G(t) = \{g\in G:\; g^{-1}t g = t\}\;\;.
\ee
One has 
\be
S \supseteq T_G\;\;,
\ee
with equality if $t$ is generic (regular). In the non-generic case, $S$ 
will be a product of simple factors and $U(1)$s.
We use the convention that $T_S=T_G$, and likewise at the Lie algebra level, 
\be
T_S=T_G\quad,\quad \lt_S = \lt_G\;\;.
\ee
At the Lie algebra level one has the (reductive, because $\lg$ is compact) decomposition
\be
\lg = \ls \oplus \lm\quad\text{with}\quad [\ls,\lm]\subset \lm \;\;.
\ee
\item 
Now let us turn to the second entry in \eqref{Adgroup}, 
\be
w \mapsto \Ad_{g^{-1}}(w + \Ad_h v - v)\;\;.
\ee
With $h=t\in T_G$ and thus $g=s\in S$, this becomes
\be
w \mapsto \Ad_{s^{-1}}(w + \Ad_t v - v)\;\;.
\ee
Decomposing $v = v^\ls + v^\lm$
into its components in $\lg = \ls \oplus \lm$, we see that only
the $\lm$-component of $v$  contributes, since $\Ad_t v^\ls = v^\ls$, 
\be
\Ad_t v - v = (\Ad_t -1) v^\lm \in \lm\;\;,
\ee
and $v^\lm$ can be chosen to cancel the $\lm$-component $w^\lm$ of $w$ 
(because by definition of $\lm$
the operator $(\Ad_t -1)$ is invertible on $\lm$). 
\item 
We are then left with 
\be
w \mapsto \Ad_{s^{-1}}w^\ls\;\;,
\ee
and we can now choose $s \in S$ such that 
\be
w \mapsto \Ad_{s^{-1}}w^\ls = \tau_2 \in \lt_S = \lt_G\;\;.
\ee
\item
Altogether, we have thus managed to conjugate 
\be
(h,w) \mapsto (t,\tau_2) \in T_G \times \lt_G\;\;,
\ee
as announced.

\end{itemize}

\item Diagonalisation / Abelianisation in $\ltg$

\begin{itemize}

\item 
The inverse adjoint action of $TG$ on its Lie algebra is
\be
\label{Ad}
(g,v)^{-1}(y,w)(g,v) = (\Ad_{g^{-1}}y,\Ad_{g^{-1}}(w + [y,v]))
\ee
In particular, for the action of $(g,0)$ one finds
\be
(g,0):\quad y \mapsto \Ad_{g^{-1}}y\quad,\quad w \mapsto \Ad_{g^{-1}}w
\ee
and for that of $(e,v)$ one has
\be
(e,v):\quad y \mapsto y\quad,\quad w \mapsto w + [y,v]\;\;.
\ee

\item First of all, we can and will choose $g\in G$ such that 
\be
y \mapsto g^{-1}yg=\tau_1\in \lt_G\;\;.
\ee
\item 
Next we define $S$ to be the stabiliser of $\tau_1$ under the adjoint action of $G$, 
\be
S:=\mathsf{Stab}_G(\tau_1) = \{g\in G:\; g^{-1}\tau_1 g = \tau_1\}\;\;.
\ee
One has 
\be
S \supseteq T_G\;\;,
\ee
with equality if $\tau_1$ is generic (regular). As above, we set
$T_S=T_G$ and $\lt_S = \lt_G$,  with
$\lg = \ls \oplus \lm$ and  $[\ls,\lm]\subset \lm$. 
\item 
Now let us turn to the second entry in \eqref{Ad},
\be
w \mapsto \Ad_{g^{-1}}(w + [y,v])
\ee
With $y=\tau_1 \in \lt_G$, and thus $g=s\in S$, this becomes
\be
w \mapsto \Ad_{s^{-1}}(w + [\tau_1,v])
\ee
Decomposing $v = v^\ls + v^\lm$, only $v^\lm$ contributes to the commutator, 
\be
[\tau_1,v] = [\tau_1,v^\lm] \in \lm\;\;,
\ee
and $v^\lm$ can be chosen to cancel $w^\lm$ (because 
by definition of $\lm$ the operator $\ad_{\tau_1}$ is invertible on $\lm$).
\item 
We are then left with 
\be
w \mapsto \Ad_{s^{-1}}w^\ls\;\;,
\ee
and as in the proof above we can now choose $s \in S$ such that 
\be
w \mapsto  \Ad_{s^{-1}}w^\ls = \tau_2 \in \lt_S = \lt_G\;\;.
\ee
\item
Altogether, we have thus managed to conjugate 
\be
(y,w) \mapsto (\tau_1,\tau_2) \in \lt_G\oplus \lt_G\;\;,
\ee
as announced.

\end{itemize}

\een

Remarks:
\ben
\item 
We see that for any $(h,w)\in TG$ or $(y,w)\in \ltg$ 
\begin{itemize}
\item one needs $g\in G/S$ (meaning $g \in G$ modulo elements in $S$) 
to conjugate $h$ or $y$ into $T_G$
\item one needs $v^m\in \lm \subset \lg$ to map $w \mapsto w^\ls \in \ls$
\item one then needs $s \in S/T_G$ to conjugate $w^\ls \mapsto \tau_2 \in \lt_G$
\end{itemize}
Thus in total, for any $g$ or $y$ (regardless of regular or not) 
the parameters $(g,v)$ lying in $TT_G \simeq T_G \times \lt_G \subset TG$ are not
needed to accomplish the Abelianisation / diagonalisation.
\item In particular, in the generic regular case ($g$ or $y$ regular), one has $S=T_G$ and
$\ls = \lt_G$. Thus the last step in the proof (consisting of conjugating $w^\ls$ into
$\lt_S = \lt_G$) is empty. The entire ambiguity in the construction then
lies in the first step, the conjugation of $h$ into $T_G$ (or $y$ into $\lt_G$). 
The possible choices of $t$ or $\tau_1$ are related by the action of the Weyl Group 
$W$, 
\be
W = N_G(T_G)/ T_G\;\;,
\ee
where $N_G(T_G)$ is the normaliser of $T_G$ in $G$. 
\item
In the opposite extreme case of $g=e$ or $y=0$, say, evidently the first step of the argument 
is empty, the stabiliser of $g=e$ or $y=0$ is of course all of $G$, 
$S=G$, and then in the last step there is then a 
$W$-fold degeneracy in the choice $\tau_2 \in \lt_G$. 
\een

\section{A Symmetry of Complex Chern-Simons Theory on Seifert 3-Manifolds}
\label{appb}

Here we establish that the symmetry that we need in section 
\ref{ab-sec} to pass from having
a background field and summing over integers at each orbifold point to
having a background field with just one integer summation to perform
exists at the level of Complex Chern-Simons theory. $BF$ theory is
then just a special limit.\footnote{It is straightforward to show that the
  symmetry is available directly in $BF$ theory but we wanted to
  establish the more general result here.} One, possibly surprising,
outcome is that the two approaches do not agree at the level of
holomorphic factorisation. 

We should also point out that these symmetries are a variation of
those found by Lawrence and Rozansky \cite{LR} and  by Mari\~{n}o
\cite{Marino}. However, here the symmetries have a clear geometric meaning
which we explain. 

\subsection{Discrete Symmetries}

The exponent in the finite dimensional integral that one arrives at in the case of
complex Chern-Simons theory is
\begin{align}
 I(k,s)&= -\frac{k}{4\pi} \Tr(\phi^{2}
 -\lambda^{2})c_{1}\left(\mathcal{L}_{M}\right) + k \sum_{1}^{N}\frac{1}{a_{i}}
  \Tr (-i \phi 
\mathbf{n}_{i} + r_{i}\pi \mathbf{n}_{i}^{2}) + k \Tr(-i\phi
\mathbf{n}_{0}) \nonumber\\ 
&   + \frac{s}{2\pi} \Tr\Big( \lambda \left(\phi
 c_{1}\left(\mathcal{L}_{M}\right)
 +2\pi i\sum_{i=1}^{N}\frac{1}{a_{i}}
\mathbf{n}_{i} + 2\pi i  \mathbf{n}_{0}\right) \Big)\label{action-fl}
\end{align}
(just use this instead of the exponent in (\ref{BF-Seifert-3})).

 The exponential of this action as well as
 the complex Ray-Singer torsion enjoy a number of  
 symmetries. We will present them here, but before doing that we note
 that the transformations below do not depend on the coupling
 constants $k$ and $s$ and so those parts of the action are
 independently invariant. Also one can
 see that the $k$-dependent part of 
 the action (\ref{action-fl}), apart from the $\lambda^{2}$ term, is
 just what one gets from Chern-Simons theory on $M$ with compact gauge
 group $G$. Furthermore, the field $\lambda$ does not transform, 
 so that the $k$-dependent part has the same invariance as for the $G$
 Chern-Simons theory.

First, we have the symmetry
\be
\mathbf{n}_{i} \rightarrow \mathbf{n}_{i} + a_{i} \mathbf{m}_{i}, \;\;
\mathbf{n}_{0} \rightarrow \mathbf{n}_{0} -\sum_{i=1}^{N} \mathbf{m}_{i}\label{shift-a}
\ee
It is clear that (\ref{action-fl}) is not changed by these transformations
(and it is consistent with our limit on the range of summation over the $\mathbf{n}_{i}$).

{\bf Remark:} Note that geometrically this symmetry is the statement
that, as the
appropriate powers of the V-line bundle $\mathcal{L}_{i}$ of
degree $1/a_{i}$ at the $i$'th orbifold point is an honest line bundle
\be
\mathcal{L}_{i}^{\otimes m_{i}a_{i}} \simeq \mathcal{L}_{0}^{\otimes m_{i}}
\ee
any orbifold line bundle $\mathcal{L}$ satisfies
\begin{align}
\mathcal{L} &= \mathcal{L}_{0}^{\otimes n_{0}} \otimes
\mathcal{L}_{1}^{\otimes n_{1}} \otimes \dots \otimes \mathcal{L}_{N}^{\otimes n_{N}}
\nonumber\\
&\simeq \mathcal{L}_{0}^{\otimes (n_{0}-\sum_{i}m_{i})} \otimes
\mathcal{L}_{1}^{\otimes (n_{1}+a_{1}m_{1})} \otimes \dots \otimes
\mathcal{L}_{N}^{\otimes (n_{N}+a_{N}m_{N}) }
\end{align}
so one has not changed the bundle but just expressed it in a
different way.

Second, we also have the transformations
\be
\mathbf{n}_{i} \rightarrow \mathbf{n}_{i} + b_{i} \mathbf{v}, \;\;
\phi \rightarrow \phi - 2\pi i
\mathbf{v},\;\;\; 
\mathbf{n}_{0} \rightarrow \mathbf{n}_{0} +b_{0}\mathbf{v}\label{shift-b}
\ee
which form a symmetry of the exponential of
the action.

{\bf Remark:} Geometrically this says that the pair $(\mathcal{L}, \,
\phi)$ of an orbifold bundle
$\mathcal{L}$ with curvature $\phi\, d\kappa$ satisfy
\be
(\mathcal{L}, \, \phi ) \simeq (\mathcal{L}\otimes
\mathcal{L}_{M}^{\otimes v}, \, \phi -2\pi i v)
\ee
The first Chern class of $\mathcal{L}_{M}^{\otimes v}$ is
\be
c_{1}(\mathcal{L}_{M}^{\otimes v}) = v.c_{1}(\mathcal{L}_{M}) 
\ee
while
\be
\frac{i}{2\pi} .\int_{\Sigma_{V}} -2\pi i v\, d\kappa = v. 
\int_{\Sigma_{V}}d\kappa = - v. c_{1}(\mathcal{L}_{M})
\ee

It is now possible to combine the symmetries and consider
\be
\mathbf{n}_{i} \rightarrow \mathbf{n}_{i} +a_{i} \mathbf{u} + b_{i} \mathbf{v}, \;\;
\phi \rightarrow \phi - 2\pi i \mathbf{v}, \;\;
\mathbf{n}_{0} \rightarrow \mathbf{n}_{0} +b_{0}\mathbf{v}- N\mathbf{u}\label{shift-c}
\ee

One then has a constructive proof that this symmetry allows one to set
all the $\mathbf{n}_{i}$ for $i\neq 0$ to zero. Recall that if the $\gcd(a, \, b)
=1$ then any integer $n$ can be expressed as
\be
n=au + bv \label{sums}
\ee
for some integers $u$. Furthermore, if $\gcd{(a, \, b)}
=1$ and $\gcd{(a, \, c)} =1$ then $\gcd{(a, \, bc)}= \gcd{(a, \,
  b)}.\gcd{(a, \, c)}=1$.

As a first step let $(\mathbf{u}, \, \mathbf{v})=(\mathbf{u}_{1}, \,
\mathbf{v}_{1})$ so that 
\be
\mathbf{n}_{1} + a_{1} \mathbf{u}_{1} + b_{1}
\mathbf{v}_{1} =0
\ee
which is guaranteed to have a solution by (\ref{sums}). All the other
$\mathbf{n}_{i}$ are changed by this but are brought back into
their appropriate ranges by the use of (\ref{shift-a}). Now we want to use
(\ref{shift-c}) again to set $\mathbf{n}_{2}=0$ while keeping
$\mathbf{n}_{1}=0$.  If the transformation 
$\mathbf{v}$ is proportional to $a_{1}$
\begin{align}
\mathbf{n}_{1} & \rightarrow 0 + a_{1}\mathbf{u}_{2} +
b_{1}a_{1}\mathbf{v}_{2} \nonumber \\
\mathbf{n}_{2} & \rightarrow \mathbf{n}_{2} +a_{2}\mathbf{u}_{2} +
b_{2}a_{1}\mathbf{v}_{2} \nonumber \\
\mathbf{n}_{3} & \rightarrow  \mathbf{n}_{3} + a_{3}\mathbf{u}_{2} +
b_{3}a_{1}\mathbf{v}_{2} \nonumber \\
\dots &  \dots \;\;\;\; \dots \;\;\; \dots\nonumber \\
\mathbf{n}_{N} & \rightarrow  \mathbf{n}_{N} + a_{N}\mathbf{u}_{2} +
b_{N}a_{1}\mathbf{v}_{2}
\end{align}
then the change in $\mathbf{n}_{1}$ is proportional to $a_{1}$
and by (\ref{shift-a}) is zero. We note
that $\gcd(a_{2},\, b_{2}a_{1}) =1$ so by (\ref{sums}) we can choose
$(\mathbf{u}_{2}, \, \mathbf{v}_{2})$ so the $\mathbf{n}_{2}$ maps to
zero. Now we would like to set $\mathbf{n}_{3}=0$ without changing
$\mathbf{n}_{1}$ and $\mathbf{n}_{2}$. To not change $\mathbf{n}_{1}$
the vector $\mathbf{v}$ must be proportional to $a_{1}$, as we saw
before, and not to change $\mathbf{n}_{2}$ it must also be proportional to
$a_{2}$ so we perform a transformation
\begin{align}
\mathbf{n}_{1} & \rightarrow 0 + a_{1}\mathbf{u}_{3} +
b_{1}a_{1}a_{2}\mathbf{v}_{3} \simeq 0\nonumber \\
\mathbf{n}_{2} & \rightarrow 0 + a_{2}\mathbf{u}_{3} +
b_{2}a_{1}a_{2}\mathbf{v}_{3} \simeq 0\nonumber \\
\mathbf{n}_{3} & \rightarrow  \mathbf{n}_{3} + a_{3}\mathbf{u}_{3} +
b_{3}a_{1}a_{2}\mathbf{v}_{3} \nonumber \\
\dots &  \dots \;\;\;\; \dots \;\;\; \dots \quad \dots\nonumber \\
\mathbf{n}_{N} & \rightarrow  \mathbf{n}_{N} + a_{N}\mathbf{u}_{2} +
b_{N}a_{1}a_{2}\mathbf{v}_{2}
\end{align}
Now $\gcd(a_{3}, \, b_{3}a_{1}a_{2})=1 $ so we can set
$\mathbf{n}_{3}=0$ by a suitable choice of $(\mathbf{u}_{3}, \,
\mathbf{v}_{3})$.

Clearly the procedure can be repeated till we have set all the
$\mathbf{n}_{i}$ to zero and the parameters we use are
\be
\mathbf{u} = \sum_{i=1}^{N}\mathbf{u}_{i}, \;\;\; \mathbf{v} =
\mathbf{v}_{1} + a_{1} \mathbf{v}_{2} + \dots + a_{1}\dots
a_{N-1}\mathbf{v}_{N}
\ee

\subsection{Holomorphic Factorisation}

The finite dimensional action for complex Chern-Simons theory
(\ref{action-fl}) can be expressed in terms of complex fields $\phi +
i \lambda$ and a complex coupling constant $t=k+is$ as 
\begin{align}
 I(t, \bar{t})&= -\frac{t}{8\pi} c_{1}(\mathcal{L}_{M}) \Tr \Phi^{2} -i
 \frac{t}{2} \Tr (\Phi \widehat{\mathbf{q}}) 
 + \frac{t\pi}{2}\sum_{i=1}^{N} \Tr ( \mathbf{n}_{i}^{2}) \\ \nonumber
 &- \frac{\bar{{t}}}{8\pi} c_{1}(\mathcal{L}_{M})\Tr \bar{\Phi}^{2} 
 + i\frac{\bar{t}}{2} \Tr (\bar{\Phi} \widehat{\mathbf{q}}) 
 + \frac{\bar{t}\pi}{2}\sum_{i=1}^{N} \Tr ( \mathbf{n}_{i}^{2})\nonumber
 \end{align}

Now we note that the holomorphic part of the action
\be
I(t) = -\frac{t}{8\pi} c_{1}(\mathcal{L}_{M}) \Tr \Phi^{2} -i
 \frac{t}{2} \Tr (\Phi \widehat{\mathbf{q}}) 
 + \frac{t\pi}{2}\sum_{i=1}^{N} \Tr ( \mathbf{n}_{i}^{2}) \label{hol-act}
 \ee
 is not invariant under the first symmetry namely under
 \be
 \mathbf{n}_{i} \rightarrow \mathbf{n}_{i} + a_{i} \mathbf{m}_{i}, \;\;
\mathbf{n}_{0} \rightarrow \mathbf{n}_{0} -\sum_{i=1}^{N} \mathbf{m}_{i}
\ee
even though both $\Phi$ and $\widehat{\mathbf{q}}$ are invariant. The quadratic term
$\mathbf{n}_{i}^{2}$ is not invariant. Furthermore, the exponential
$\exp{\left(i I(t)\right)}$ is also not invariant, unless $s=0$, as the coupling
constant is complex and so one does not just get a phase.

This implies that we can have two inequivalent
holomorphic factorizations. The first is to take the partition function
with all the $\mathbf{n}_{i}$ switched on and then factorise with
holomorphic action (\ref{hol-act}) while the second is to set all the
$\mathbf{n}_{i}=0$ for $i \neq 0$ and then get the factorisation in
\cite{BT-CCS}.

\rnc{\Large}{\normalsize}

\end{document}